\documentclass[twocolumn,twocolappendix]{aastex631}

\usepackage{url}
\usepackage{amsmath}

\newcommand\tdasalt{$42.8_{-9.7}^{+11.4}$\,d}
\newcommand\tdbsalt{$5.6_{-2.1}^{+2.3}$\,d}
\newcommand\tdcsalt{$26.7_{-4.2}^{+2.5}$\,d}
\newcommand\tdabsalt{$-103.6_{-32.2}^{+27.9}$\,d}
\newcommand\tdbcsalt{$57.7_{-12.3}^{+10.2}$\,d}
\newcommand\tdacsalt{$-46.8_{-33.7}^{+29.5}$\,d}

\newcommand\muasalt{$9.3_{-3.7}^{+5.2}\,$}
\newcommand\mubsalt{$16.2_{-3.2}^{+11.3}\,$}
\newcommand\mucsalt{$10.3_{-2.0}^{+1.7}\,$}
\newcommand\muaysalt{$9.0_{-3.6}^{+4.8}\,$}
\newcommand\mubysalt{$15.6_{-3.0}^{+11.1}\,$}
\newcommand\mucysalt{$9.8_{-2.0}^{+1.6}\,$}
\newcommand\tdahsiao{$50.6_{-15.3}^{+16.1}$\,d}
\newcommand\tdbhsiao{$6.5_{-1.8}^{+2.4}$\,d}
\newcommand\tdchsiao{$24.3_{-3.9}^{+3.9}$\,d}
\newcommand\tdabhsiao{$-122.3_{-43.8}^{+43.7}$\,d}
\newcommand\tdbchsiao{$49.3_{-14.7}^{+12.2}$\,d}
\newcommand\tdachsiao{$-73.6_{-46.1}^{+44.7}$\,d}

\newcommand\muahsiao{$12.6_{-5.9}^{+9.9}\,$}
\newcommand\mubhsiao{$15.2_{-2.7}^{+8.6}\,$}
\newcommand\muchsiao{$8.2_{-1.9}^{+1.8}\,$}
\newcommand\tbsfa{6.2}
\newcommand\tbsfb{9.6}
\newcommand\tbsfc{0.4}
\newcommand\tbsfd{0.0}
\newcommand\tbsfe{2.3}
\newcommand\tcsfa{28.6}
\newcommand\tcsfb{41.3}
\newcommand\tcsfc{38.6}
\newcommand\tcsfd{30.3}
\newcommand\tcsfe{40.0}
\newcommand\tabtam{{45.1}}
\newcommand\tabtal{{-8.2}}
\newcommand\tabtau{{+6.1}}
\newcommand\tabtbm{{4.2}}
\newcommand\tabtbl{{-1.0}}
\newcommand\tabtbu{{+1.1}}
\newcommand\tabtcm{{27.5}}
\newcommand\tabtcl{{-0.7}}
\newcommand\tabtcu{{+0.8}}

\newcommand\tabtamind{{36.9}}
\newcommand\tabtalind{{-9.0}}
\newcommand\tabtauind{{+7.7}}
\newcommand\tabtbmind{{4.9}}
\newcommand\tabtblind{{-1.1}}
\newcommand\tabtbuind{{+0.7}}
\newcommand\tabtcmind{{27.8}}
\newcommand\tabtclind{{-0.8}}
\newcommand\tabtcuind{{+2.2}}
\newcommand\tabtamhsiaojt{{52.5}}
\newcommand\tabtalhsiaojt{{-11.4}}
\newcommand\tabtauhsiaojt{{+11.0}}
\newcommand\tabtbmhsiaojt{{5.6}}
\newcommand\tabtblhsiaojt{{-0.4}}
\newcommand\tabtbuhsiaojt{{+0.4}}
\newcommand\tabtcmhsiaojt{{26.0}}
\newcommand\tabtclhsiaojt{{-0.8}}
\newcommand\tabtcuhsiaojt{{+1.0}}
\newcommand\tabtamhsiao{{43.4}}
\newcommand\tabtalhsiao{{-22.4}}
\newcommand\tabtauhsiao{{+22.2}}
\newcommand\tabtbmhsiao{{5.9}}
\newcommand\tabtblhsiao{{-0.6}}
\newcommand\tabtbuhsiao{{+0.5}}
\newcommand\tabtcmhsiao{{26.9}}
\newcommand\tabtclhsiao{{-1.0}}
\newcommand\tabtcuhsiao{{+1.1}}
\newcommand\tabtbmsnida{0.5}
\newcommand\tabtbesnida{4.2}
\newcommand\tabtcmsnida{25.8}
\newcommand\tabtcesnida{13.7}
\newcommand\tabtbmsnidb{5.0}
\newcommand\tabtbesnidb{6.8}
\newcommand\tabtcmsnidb{26.0}
\newcommand\tabtcesnidb{12.6}

\newcommand\ttbsnidsubtype{$0.3_{-4.3}^{+4.3}$}
\newcommand\ttcsnidsubtype{$23.3_{-13.3}^{+13.3}$}

\newcommand{\kms}{{km}\,{s}$^{-1}$}

\newcommand{\brenda}[1]{\textcolor{black}{#1}}

\newcommand{\spw}[1]{\textcolor{black}{#1}}

\begin{document}



\title{JWST Spectroscopy of SN H0pe: Classification and Time Delays of a Triply-imaged Type Ia Supernova at $z=1.78$}

\correspondingauthor{Wenlei Chen}
\email{wenlei.chen@okstate.edu}

\author[0000-0003-1060-0723]{Wenlei Chen}
\affiliation{Department of Physics, Oklahoma State University, 145 Physical Sciences Bldg, Stillwater, OK 74078, USA}
\affiliation{School of Physics and Astronomy, University of Minnesota, 116 Church Street SE, Minneapolis, MN 55455, USA}

\author[0000-0003-3142-997X]{Patrick L. Kelly}
\affiliation{School of Physics and Astronomy, University of Minnesota, 116 Church Street SE, Minneapolis, MN 55455, USA}

\author[0000-0003-1625-8009]{Brenda L.~Frye}
\affiliation{Department of Astronomy/Steward Observatory, University of Arizona, 933 N. Cherry Avenue, Tucson, AZ 85721, USA}

\author[0000-0002-2361-7201]{Justin Pierel}
\affiliation{Space Telescope Science Institute, 3700 San Martin Dr., Baltimore, MD 21218, USA}

\author[0000-0002-9895-5758]{S.\ P.\ Willner}
\affiliation{Center for Astrophysics \textbar\ Harvard \& Smithsonian, 60 Garden Street, Cambridge, MA, 02138, USA}

\author[0000-0002-2282-8795]{Massimo Pascale}
\affiliation{Department of Astronomy, University of California, 501 Campbell Hall \#3411, Berkeley, CA 94720, USA}

\author[0000-0003-3329-1337]{Seth H. Cohen} 
\affiliation{School of Earth and Space Exploration, Arizona State University,
Tempe, AZ 85287-1404, USA}

\author[0000-0003-1949-7638]{Christopher J. Conselice} 
\affiliation{Jodrell Bank Centre for Astrophysics, Alan Turing Building,
University of Manchester, Oxford Road, Manchester M13 9PL, UK}

\author{Michael Engesser}
\affiliation{Space Telescope Science Institute, 3700 San Martin Drive, Baltimore, MD 21218, USA}

\author[0000-0001-6278-032X]{Lukas J. Furtak}
\affiliation{Physics Department, Ben-Gurion University of the Negev, P. O. Box 653, Be’er-Sheva, 8410501, Israel}

\author[0000-0002-5116-7287]{Daniel Gilman} 
\affiliation{Department of Astronomy $\&$ Astrophysics, University of Chicago, Chicago, IL 60637, USA }
\affiliation{Brinson Fellow}

\author[0000-0001-9440-8872]{Norman A. Grogin}
\affiliation{Space Telescope Science Institute, 3700 San Martin Dr., Baltimore, MD 21218, USA}

\author[0000-0002-6741-983X]{Simon Huber} 
\affiliation{Max-Planck-Institut f\"ur Astrophysik, Karl-Schwarzschild-Str. 1, 85748 Garching, Germany}
\affiliation{Physik-Department, Technische Universitaet Muenchen, James-Franck-Straße 1, 85748 Garching, Germany}

\author[0000-0001-8738-6011]{Saurabh W. Jha}
\affiliation{Department of Physics and Astronomy, Rutgers, the State University of New Jersey, 136 Frelinghuysen Rd., Piscataway, NJ 08854, USA}

\author[0000-0001-5975-290X]{Joel Johansson}
\affiliation{Oskar Klein Centre, Department of Physics, Stockholm University, AlbaNova, SE-10691 Stockholm, Sweden}

\author[0000-0002-6610-2048]{Anton M. Koekemoer}
\affiliation{Space Telescope Science Institute, 3700 San Martin Dr., Baltimore, MD 21218, USA}

\author[0000-0003-2037-4619]{Conor Larison}
\affiliation{Department of Physics and Astronomy, Rutgers, the State University of New Jersey, 136 Frelinghuysen Rd., Piscataway, NJ 08854, USA}

\author[0000-0002-7876-4321]{Ashish K. Meena}
\affiliation{Physics Department, Ben-Gurion University of the Negev, P. O. Box 653, Be’er-Sheva, 8410501, Israel}

\author[0000-0003-2445-3891]{Matthew R. Siebert}
\affiliation{Space Telescope Science Institute, 3700 San Martin Dr., Baltimore, MD 21218, USA}

\author[0000-0001-8156-6281]{Rogier A. Windhorst} 
\affiliation{School of Earth and Space Exploration, Arizona State University,
Tempe, AZ 85287-1404, USA}

\author[0000-0001-7592-7714]{Haojing Yan}
\affiliation{Department of Physics and Astronomy, University of Missouri,
Columbia, MO 65211, USA}

\author[0000-0002-0350-4488]{Adi Zitrin}
\affiliation{Physics Department, Ben-Gurion University of the Negev, P. O. Box 653, Be’er-Sheva, 8410501, Israel}



\begin{abstract}
SN H0pe is a triply imaged supernova (SN) at redshift $z=1.78$ discovered using the {\it James Webb Space Telescope} {\it (JWST)}. 
In order to classify the SN spectroscopically and measure the relative time delays of its three images (designated A, B, and C), we acquired NIRSpec follow-up spectroscopy spanning 0.6 to 5~\micron. From the high signal-to-noise spectra of the two bright images B and C, we first classify the SN, whose spectra most closely match those of SN 1994D and SN 2013dy, as a Type Ia SN. We identify prominent blueshifted absorption features corresponding to \ion{Si}{2}\,$\lambda6355$ and \ion{Ca}{2} H\,$\lambda3970$ and K\,$\lambda3935$. We next measure the absolute phases of the three images from our spectra, which allows us to constrain their relative time delays. The absolute phases of the three images,  determined by fitting the three spectra to Hsiao07 SN templates, are \tdbhsiao, \tdchsiao\, and \tdahsiao\ for the brightest to faintest images. These correspond to relative time delays between Image A and Image B and between Image B and Image C of \tdabhsiao\, and \tdbchsiao, respectively. The SALT3-NIR model yields phases and time delays consistent with these values. After unblinding, we additionally explored the effect of using Hsiao07 template spectra for simulations through eighty instead of sixty days past maximum, and found a small (11.5 and 1.0 days, respectively) yet statistically insignificant ($\sim$0.25$\sigma$ and $\sim$0.1$\sigma$) effect on the inferred image delays.
\end{abstract}



\section{Introduction} \label{sec:intro}

\citet{refsdal64} showed that, in principle, the relative time delays between the appearances of a multiply imaged supernova (SN) could be used to constrain the Hubble constant $H_0$. It took more than fifty years, however, to put this theory to use. The first strongly lensed SN with multiple resolved images, dubbed ``SN Refsdal'' in honor of Sjur Refsdal,  was discovered in 2014 \citep{kellyrodneytreu15,kellyrodneytreu2016}, and it has been used to measure $H_0$ with competitive precision \citep{kellyrodneytreu2023b}. SN Refsdal occurred in a galaxy at $z=1.491$ \citep{grahamebelinglimousin09} and was strongly lensed by the MACS~J1149.5+2223 galaxy cluster.
The method of measuring $H_0$ using a SN strongly lensed by a galaxy cluster involves systematic uncertainties that are different from those of  measurements that instead employ quasars that are multiply imaged by galaxy-scale lenses \citep[e.g.,][]{2017MNRAS.468.2590S}.
In particular, models of galaxy-cluster lenses have substantially less sensitivity to the mass-sheet degeneracy given strongly lensed sources at multiple redshifts, and they exhibit more modest projection effects in comparison to galaxy-scale lenses.
Being lensed by a galaxy cluster, SN Refsdal also had an approximately year-long time delay, which is much longer than those from galaxy-scale lenses. This enabled more precise time-delay measurements to constrain $H_0$.
Based on a preliminary time delay measurement of its multiple images from \citet{kellyrodneytreu2016} and  available model predictions, \citet{vegaferrero2018} obtained $H_0=64_{-11}^{+9}$~\kms~{Mpc}$^{-1}$ at 68\% confidence. With a 1.5\% time-delay measurement from \citet{kellyrodneytreu2023b}, \citet{kellyrodneytreu2023a} performed a blind measurement of the relative time delays and magnification ratios of its multiple images with improved precision and accuracy. Based on eight independent cluster lens models, \citet{kellyrodneytreu2023b} inferred $H_0=64.8_{-4.3}^{+4.4}$~\kms~{Mpc}$^{-1}$.  The two lens models most consistent with the observations yielded $H_0=66.6_{-3.3}^{+4.1}$~\kms~{Mpc}$^{-1}$. 
Recent efforts have used the 1.5\% \citet{kellyrodneytreu2023a} time delay measurement for Refsdal in combination with additional lens models to constrain the Hubble constant for additional assumptions, and obtain measurements within $\sim$1-$\sigma$ agreement \citep{2024arXiv240213476L,2024arXiv240110980G}. 

\begin{figure*}
\centering
\includegraphics[angle=0,width=6in]{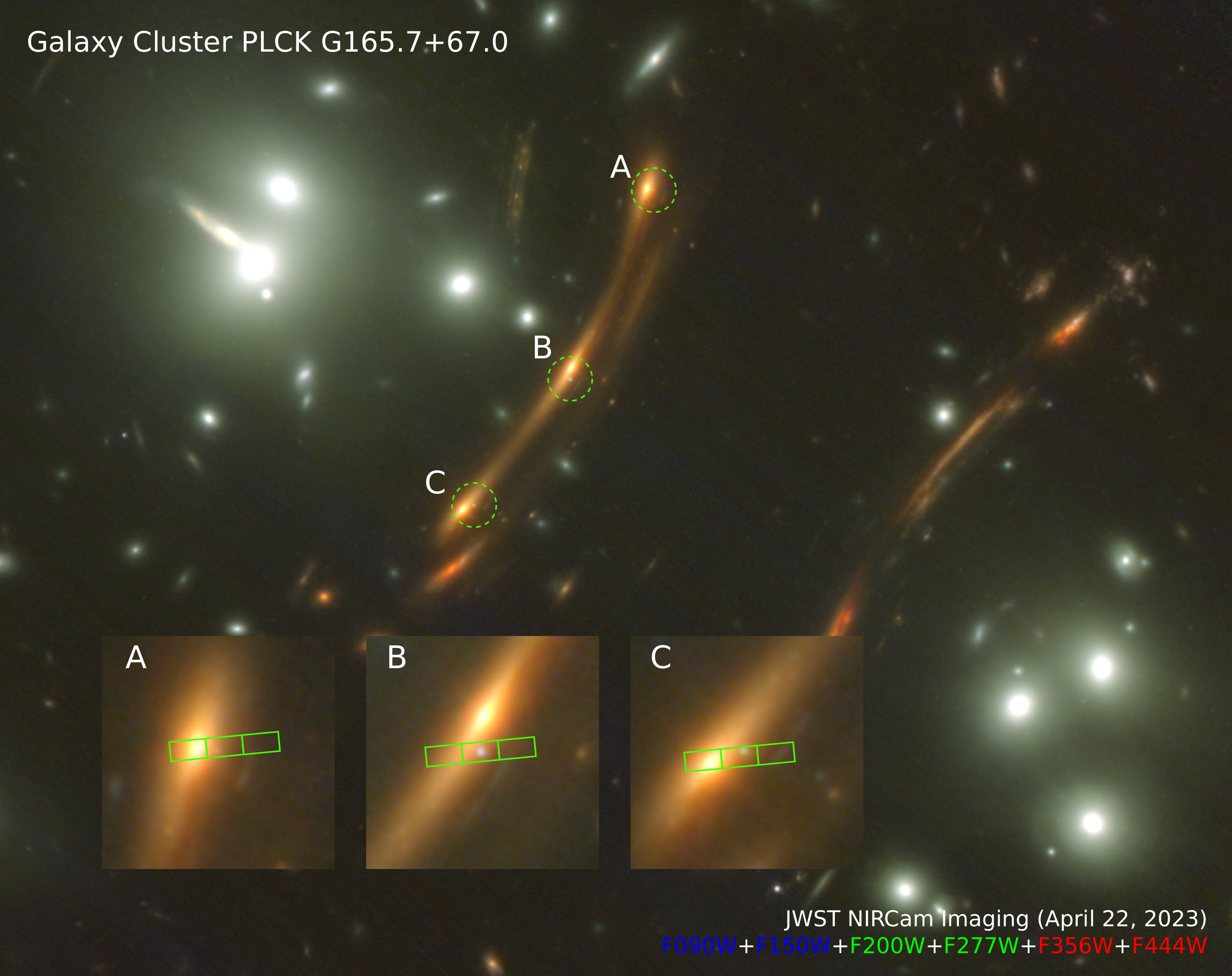}
\spw{\caption{Color image  of the Arc~2 region of G165. This ``Epoch~2'' NIRCam image was taken on 2023 April 22, the same day the spectra were acquired. Green dashed circles in the main image are centered on the three SN images, which are labeled A, B, C from north to south.  The three stamp images superposed on the main image show magnified images of the SN and its host galaxy with superposed green rectangles marking the open shutters in the NIRSpec MSA\null. The SN is visible as a blue point source situated at the central shutter of each slitlet. North is up and east is to the left in all images, and red represents the NIRCam F356W and F444W filters, green represents F200W and F277W, and blue represents F090W and F150W\null.  The scale of the main image is indicated by the distance from A to C, which is 14\arcsec.}}
\label{fig:ddt4446}
\end{figure*}

Additional multiply imaged SNe have been found. SN~Ia iPTF~16geu at $z=0.409$ was detected in ground-based imaging two years after SN Refsdal \citep{goobaramanullahkulkarni17}.
Two other SNe, ``SN Requiem'' at $z = 1.95$ \citep{rodneybrammerpierel21} and ``SN Zwicky'' at $z=0.354$ \citep{goobarjohanssonschulze2022}, were identified in {\it Hubble Space Telescope} {\it (HST)} data. 
Recently, a multiply imaged core-collapse supernova (CCSN) at $z\approx3$ was found in archival {\it HST} images taken in 2010 \citep{chenkellyoguri2022}. In addition, a candidate  multiply imaged SN Ia (SN~2022riv) was detected in {\it HST} imaging ({\it HST} SNAP 16729; PI: P.\ Kelly) and followed up by {\it HST} and {\it James Webb Space Telescope} ({\it JWST}) observations ({\it HST} 16264; PI: J.\ Pierel, {\it HST} 17253; PI: P.\ Kelly, and {\it JWST} 2767; PI: P.\ Kelly), making it the first strongly lensed SN observed by {\it JWST}\null.
However, these cases all have limitations for cosmological constraints. SN iPTF~16geu and SN Zwicky are at relatively low redshifts, and the short time delays ($<$1 day) between
the multiple images do not allow an accurate measurement of $H_{0}$ \citep{dhawan2020,johansson2021}. SN Requiem and the CCSN at $z\approx3$ lack light curves for the measurement of the relative delays, and only the trailing image of 22riv has been detected. 

On 2023 March 30, a candidate SN Ia dubbed ``SN H0pe'' \citep{Frye2023an} appeared in the NIRCam images of the PLCK G165.7+67 (G165) galaxy-cluster field. SN H0pe was confirmed to be a transient when the NIRCam images were compared against {\it HST} WFC3 images taken \brenda{seven years} earlier. Follow-up spectroscopic observations using the Large Binocular Telescope (LBT) were also conducted, which yielded a precise redshift for the arc \citep{2023A&A...675L...4P}. At $z=1.78$ \citep{2023A&A...675L...4P,Frye_2024}, SN H0pe is the highest-redshift strongly lensed SN Ia.
 The candidate was followed up \spw{23 days} after discovery
with NIRSpec spectroscopy \spw{and a second} epoch of NIRCam imaging (Figure~\ref{fig:ddt4446}), which was followed by\spw{a third NIRCam imaging epoch  17 days later\citep{Frye_2024}.}  
SN H0pe offers the opportunity to make a second, independent measurement of $H_0$ using a different cluster lens. 

\citet{Frye_2024} described SN H0pe's discovery and \brenda{the strong lensing}, photometric, and spectroscopic analysis of the G165 cluster field. This paper analyzes the spectroscopic data to determine the SN's phase in each of its three images. Pierel et al.\ (2024, submitted) have analyzed photometry of SN H0pe in order to obtain an independent measurement the relative time delays among the images.
An upcoming publication by Pascale et al.~2024 will present an inference of $H_0$ based on the time-delay measurements from these spectroscopic and photometric analyses.
For this study, the spectroscopic phases were measured in a blind analysis independent of the light curve and the lens modeling. In other words, we intentionally excluded imaging data and lens models from the current analysis, and the results are based only on the NIRspec data and SN spectral templates. 

The paper is organized as follows. Section~\ref{sec:data} describes the {\it JWST} NIRSpec data set. Section~\ref{sec:gal_spec} and Section~\ref{sec:sn_spec} presents the spectral analysis of SN H0pe and its host galaxy. Section~\ref{sec:sn_phases} describes our methods to measure the phases of the SN images. Section~\ref{sec:method_validation} outlines the simulation of our model and the error analysis.
Section~\ref{sec:final_constraints} presents the final constraints on the time delays and the magnifications of SN H0pe's images.
Section~\ref{sec:hsiao80} describes the updates after we unblinded our initial results to the $H_0$-inference team (Pascale et al.~2024).
Section~\ref{sec:discussion} discusses the implications and presents our conclusions.

\section{Data}
\label{sec:data}

SN H0pe was detected in {\it JWST} NIRCam imaging of the galaxy cluster G165 acquired on UT 2023 March 30 by the \brenda{Prime Extragalactic Areas for Reionization and Lensing Science (PEARLS) program (\citealt{Windhorst2023}, GTO 1176; PI: R.\ Windhorst). \citet{Frye_2024} described the data reduction and analysis \spw{and the three images of the SN and its host galaxy}.} The SN and its host galaxy are near the central region of the galaxy-cluster field, as shown in Figure~\ref{fig:ddt4446}. SN Images A, B, and C have
R.A.\ and Dec.\ coordinates (J2000) 11:27:15.31, +42:28:41.0; 11:27:15.60, +42:28:33.8; and 11:27:15.94, +42:28:28.9, respectively \spw{\citep[][their Table~2]{Frye_2024}}. 

We obtained follow-up {\it JWST} NIRSpec Multi-Object Spectroscopy (MOS) and NIRCam imaging on 2023 April 22 UT (``Epoch~2:'' DDT 4446; PI: B.\ Frye).
The NIRSpec observations were designed using the Micro-Shutter Assembly (MSA) to acquire spectra of the three images of SN H0pe, two of the three images of the SN host galaxy, and 42 other gravitationally lensed arcs.
The SN and host galaxy spectra used three MSA slitlets (open shutters) end-to-end as shown in Figure~\ref{fig:ddt4446}. The color image was generated from the Epoch~2 NIRCam images.
Each slitlet has an open area of 0\farcs20 in the dispersion direction and 0\farcs46 in the spatial direction.  There is a 0\farcs07 gap between slitlets, giving a total height for three slitlets of 1\farcs52. The NIRSpec MOS data \brenda{comprised} the PRISM (0.6 to 5.3~\micron\ wavelength coverage at spectral resolution $R\sim100$) and the G140M and G235M gratings (0.70 to 3.07~\micron\ at $R\sim1000$). 



We began processing the spectroscopic data by downloading the Stage~2 data from the Mikulski Archive for Space Telescopes (MAST)\null. Additional processing used the {\it JWST} pipeline\footnote{\url{https://github.com/spacetelescope/jwst}} \citep{Bushouse2022} with context file jwst\_1087.pmap to produce two-dimensional (2D) spectral data. The pipeline applied a slit-loss throughput correction for the SN images based on their planned positions within the MSA shutters. The spectra of the SN and its host galaxy images were extracted using the optimal extraction algorithm from \citet{horne1986} implemented as scripts available as part of the MOS Optimal Spectral Extraction (MOSE) 
notebook.\footnote{\url{https://spacetelescope.github.io/jdat_notebooks/notebooks/optimal_extraction/Spectral_Extraction-static.html}} 
We used {\tt webbpsf}\footnote{\url{https://webbpsf.readthedocs.io/en/latest/index.html}} to generate the effective point-spread functions (PSFs) for the NIRSpec observations. 

\begin{figure*}
\centering
\includegraphics[angle=0,width=6in]{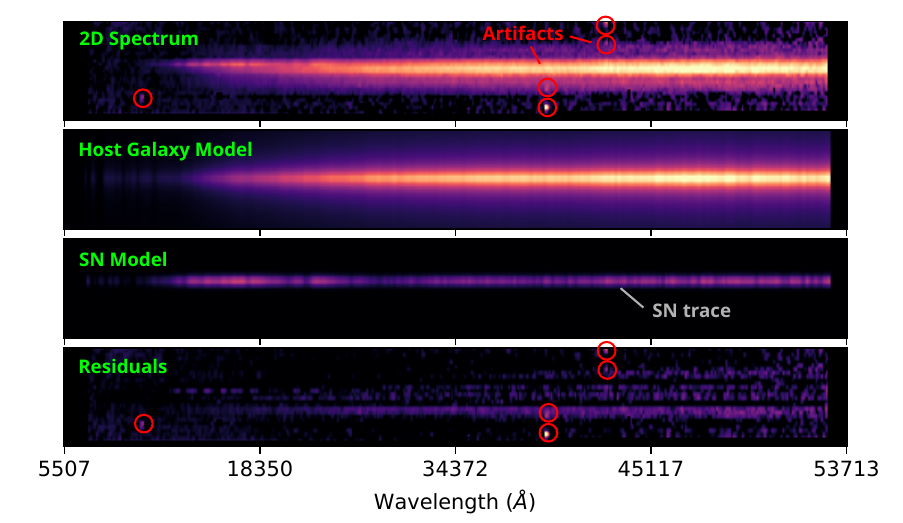}
\caption{NIRSpec PRISM spectrum \brenda{of the host galaxy and the SN} from Slit~B (Figure~\ref{fig:ddt4446}). \spw{Observed} wavelengths are marked at the bottom, and the total height of each panel projects to 1\farcs52 on the sky. The top panel shows the 2D spectrum from the {\it JWST} pipeline. The middle two panels show the best-fit models of the host galaxy and the SN, respectively. The bottom panel shows the residuals after subtracting the galaxy and the SN models from the original spectrum. Red circles label the cosmic-ray-related artifacts (hot pixels) in the 2D spectrum.}
\label{fig:2d_model}
\end{figure*}

To separate the spectra of the SN and its host galaxy, we fit two kernels simultaneously to the flux in the 2D spectrum as a function of wavelength. We used a Gaussian \spw{kernel} to model the flux distributions of the point source (the SN) and selected from Gaussian, Moffat \citep{moffat1969}, and Voigt \citep[e.g.,][]{1968JQSRT...8.1379W} distribution functions to model the extended source (the host galaxy) along the spatial direction of each 2D spectrum. The \spw{kernels} were chosen based on the best-fit profile functions that minimized the least-square statistic. During the extraction process, we disregarded the uncertainties related to the kernel functions. Figure~\ref{fig:2d_model} shows the 2D spectrum model and the extracted spectrum for the PRISM observations of Image B\null. 
Appendix~\ref{appendix:2d_spec} gives complete details of the extraction method and the full complement of two-dimensional spectra and source models \brenda{for the SN\null. 
\citet{Frye_2024}} \spw{presented the full spectroscopic data set.
}

\section{Spectrum of the host galaxy}
\label{sec:gal_spec}

\begin{figure*}
\centering
\includegraphics[angle=0,width=\linewidth]{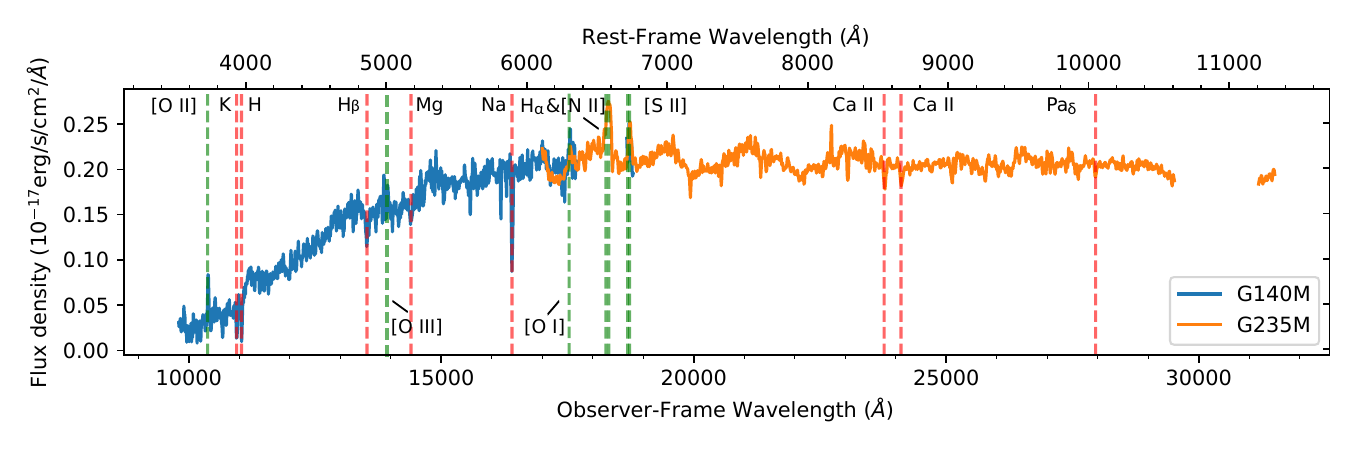}
\caption{\brenda{NIRSpec G140M and G235M spectra of the A image of the host-galaxy nucleus.} The data are for the single 0\farcs46 slitlet containing the nucleus (Figure~\ref{fig:ddt4446}). Vertical lines mark the positions of the detected emission (green) and absorption (red) lines, as listed in Table~\ref{tab:lines}. Wavelengths are marked in the observed frame (bottom) and in the $z=1.7825$ rest frame (top). The gap in the spectrum near observed 3\,\micron\ is from the physical gap between the two  NIRSpec detector chips. \brenda{The spectra and line identifications for Images A and C are also shown by \citet{Frye_2024}.}}
\label{fig:spec_z}
\end{figure*}

As shown in Figure~\ref{fig:ddt4446}, the NIRSpec slitlets placed on Images A and C intersect the host galaxy's nucleus. Because the SN was faintest in image~A, the A slitlet yields the highest signal-to-noise ratio and least-contaminated spectrum of the host galaxy.  Figure~\ref{fig:spec_z} shows the spectrum 
of the host, and Table~\ref{tab:lines} lists the identifications and fluxes of the detected lines.
For our analysis, we used {\tt GLEAM} \citep{gleam}, a software package that 
uses the {\tt LMFIT} Python package \citep{2021zndo...5570790N} to perform the line fitting and to calculate errors on fit parameters.
The line wavelengths (excluding unresolved doublets such as [\ion{O}{2}]) give spectroscopic redshift
$z=1.7825\pm0.0008$, which agrees with the spectroscopic measurement using the LBT Utility Camera in the Infrared (LUCI) \citep{2023A&A...675L...4P}
and with an independent redshift measurement \citep{Frye_2024} from the same NIRSpec data.
\brenda{The line identifications also agree with \citet{Frye_2024}, who analyzed the spectroscopy of host-galaxy Images A and C (labeling them Arc~2a and~2c, respectively). They estimated the H${\alpha}$ line flux with corrections for underlying stellar absorption, dust extinction, and slit loss of the whole arc. Their Table~4 gives host-galaxy \brenda{specific} star-formation rates (derived from the H${\alpha}$ flux) and stellar masses 
\brenda{derived from Images~A and~C.}
}

\begin{table*}
\centering
\caption{\textbf{
Emission and absorption lines of the A image of the host-galaxy nucleus.} 
}
\begin{tabular}{@{}lrrr@{}}
\hline
 \textbf{Spectral Line} & \textbf{Wavelength} & \textbf{Line Flux\tablenotemark{a}} & \textbf{EW\tablenotemark{b}} \\
  & (rest-frame \AA) & ($\times 10^{-17}$ erg s$^{-1}$ cm$^{-2}$) & (\AA) \\
\hline
{[O~\textsc{ii}]}, {[O~\textsc{ii}]} & 3727, 3730 & $1.40 \pm 0.22$ & $15.6 \pm 2.5$ \\
{Ca K} & 3935 & $-1.13 \pm 0.26$\rlap{\tablenotemark{c}} & $-8.0 \pm 1.9$ \\
{Ca H} & 3970 & $-1.86 \pm 0.40$ & $-10.5 \pm 2.3$ \\
{H${\beta}$} & 4863 & $-2.09 \pm 0.32$ & $-4.9 \pm 0.7$ \\
{[O~\textsc{iii}]} & 5008 & $0.75 \pm 0.23$ & $1.7 \pm 0.5$ \\
{Mg} & 5177 & $-1.66 \pm 0.39$ & $-3.6 \pm 0.8$ \\
{Na} & 5896 & $-4.07 \pm 0.22$ & $-7.3 \pm 0.4$ \\
{[O~\textsc{i}]} & 6302 & $1.97 \pm 0.42$ & $3.5 \pm 0.7$ \\
{H${\alpha}$}, {[N~\textsc{ii}]} & 6565, 6585 & $7.24 \pm 0.87$ & $12.1 \pm 1.5$ \\
{[S~\textsc{ii}]}, {[S~\textsc{ii}]} & 6718, 6732 & $3.39 \pm 0.37$ & $6.0 \pm 0.7$ \\
{Ca~\textsc{ii}} & 8544 & $-1.41 \pm 0.17$ & $-2.5 \pm 0.3$ \\
{Ca~\textsc{ii}} & 8665 & $-1.15 \pm 0.14$ & $-2.0 \pm 0.2$ \\
{Pa${\delta}$} & 10053 & $-0.65 \pm 0.10$ & $-1.1 \pm 0.2$ \\
\hline
\end{tabular}
\label{tab:lines}
\begin{flushleft}
\tablenotetext{a}{The flux measurements are as observed in a single slitlet, i.e., not corrected for magnification due to gravitational lensing, \brenda{underlying stellar absorption, dust extinction, or \spw{slit losses}. \citet{Frye_2024} first identified the lines listed and showed spectra of host-galaxy Images A and C (labeling them Arc~2a and~2c, respectively).  They also gave the H${\alpha}$ line flux with corrections for underlying stellar absorption, dust extinction, and slit loss of the whole arc.}}
\tablenotetext{b}{Rest-frame equivalent widths (EWs) of emission and absorption lines.}
\tablenotetext{c}{Negative flux and EWs indicate absorption lines\spw{, the latter opposite the usual convention.}}
\end{flushleft}
\end{table*}

\section{Spectra of SN H0pe}
\label{sec:sn_spec}


\begin{figure*}
\centering
\includegraphics[angle=0,width=5.8in]{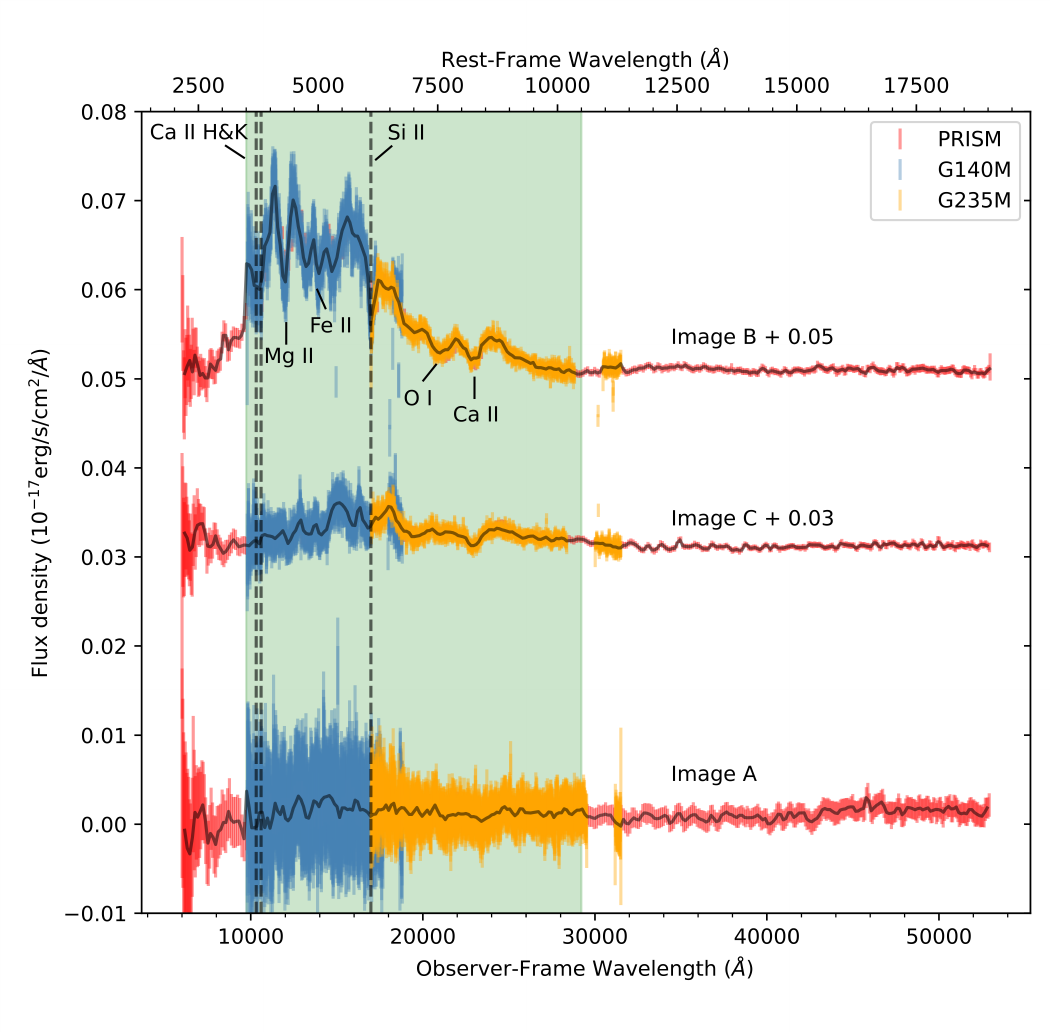}
\caption{NIRSpec spectra of all three images of SN~H0pe. Spectra are offset vertically as labeled for clarity. Error bars, merging into color bands as indicated in the legend, show the G140M, G235M, and PRISM spectra. Solid lines show the smoothed spectra using a 120\AA\ rolling window. Wavelengths are marked in the observed frame (bottom) and the rest frame (top\spw{, based on the host-galaxy $z=1.7825$}). The green shading marks the wavelength range used to match the SN templates and  measure the image phases. Dashed vertical lines mark the wavelengths of key absorption features. Spectra are plotted as extracted in a single slitlet, not corrected for slit losses.}
\label{fig:snspec_raw}
\end{figure*}

\begin{figure*}
\centering
\includegraphics[angle=0,width=6in]{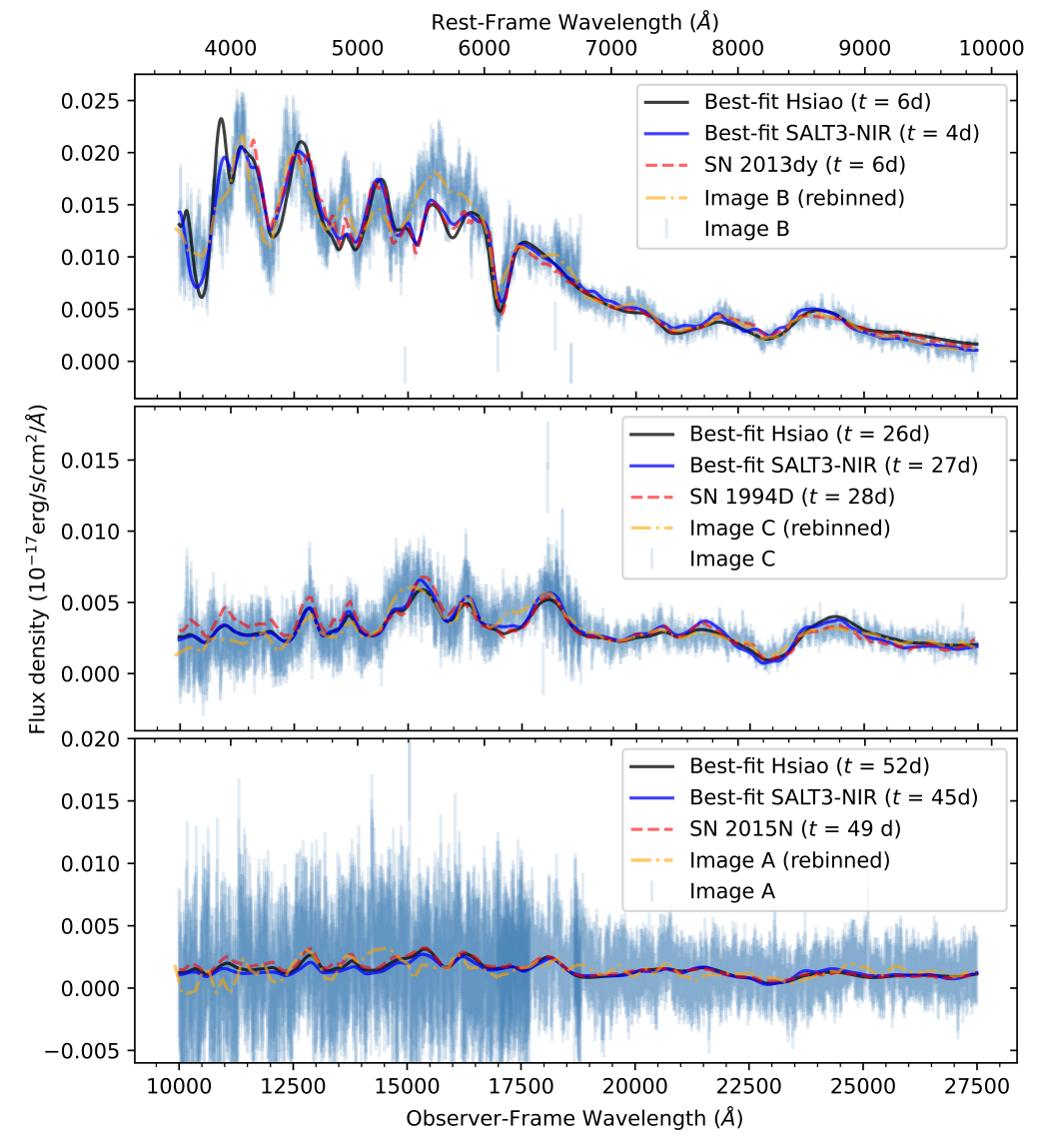}
\caption{Best-fit templates \brenda{and phases} for the three images of SN~H0pe \spw{as shown in the legends}.  
\spw{Thin blue lines, merging into color bands, show 1$\sigma$ error bars.}
Dash-dotted orange lines show the \spw{observed} spectra rebinned into a 120\,\AA\ rolling window.
Solid black lines show the best-fit Hsiao07 spectra. Solid blue lines show the best-fit SALT3-NIR model at each phase. Red dashed lines show spectra of previously observed SNe Ia that most closely match the observed spectra of SN~H0pe from our {\tt NGSF} analysis.}
\label{fig:snspec}
\end{figure*}

Figure~\ref{fig:snspec_raw} shows the extracted spectra of Images B and C\null. The portions within rest wavelengths 
3\,600\,\AA\ to 10\,000\,\AA\ were cross matched with the SN~Ia template library available in 
the \brenda{Next Generation SuperFit ({\tt NGSF})} software package \citep{ngsf}. Table~\ref{tab:superfit} lists the five best-fit SN templates from {\tt NGSF}\null. As shown in Figure~\ref{fig:snspec}, the SN spectra are consistent with those of a SN Ia. The spectra of Images B and C most closely match those SN 2013dy and SN 1994D at phases of 6 days and 28 days, respectively.
The favored subtype is ``Ia-norm,'' a normal SN Ia \citep{snid}. Image~A had the lowest signal-to-noise ratio among the three SN images, and consequently the template-matching could not yield a favored SN type or subtype. If we force the match within a set of previously observed SNe Ia, the spectrum of Image A most closely matches SN 2015N at 49 days after its peak brightness. Figure~\ref{fig:snspec} shows these best-match spectra superimposed on the spectra of SN H0pe for comparison.


When the spectra were obtained, Image B was the brightest of the three SN images. As shown in Figure~\ref{fig:snspec_raw}, the spectrum of Image B exhibits the \ion{Si}{2} absorption feature near 6120\,\AA\ in the rest frame of the host galaxy. This is an identifying characteristic of SN Ia spectra \citep{filippenko1997}: A deep absorption trough around 6150\,\AA\, produced by
blueshifted \ion{Si}{2}\,$\lambda$6355 emission 
is prominent in the spectra of SNe Ia through roughly several weeks after maximum but is absent from the spectra of other types of SNe.

Typical velocities for SNe Ia near maximum light are $v_\mathrm{Si\,II} \approx 10\hbox{--}12\times 10^{3}$~{km}\,{s}$^{-1}$ and $v_\mathrm{Ca\, II} \approx 13\hbox{--}15\times 10^{3}$~{km}\,{s}$^{-1}$ \citep{filippenko1997}. 
Figure~\ref{fig:snspec_raw} shows blueshifted \ion{Si}{2}\,$\lambda$6355 and \ion{Ca}{2} H \& K ($\lambda3935$, $\lambda3970$) absorption lines at $6117 \pm 2$\,\AA\ (${\sim} 17017$\,\AA\ in the observer frame) and at $3790 \pm 2$\,\AA\ (${\sim} 10544$\,\AA\ in the observer frame), respectively. 
These wavelengths for SN H0pe give $v_\mathrm{Si\,II}=-11.23 \pm 0.11 \times 10^{3}$~\kms\ and $v_\mathrm{Ca\,II}=-12.54 \pm 0.19 \times 10^{3}$~\kms\ relative to the host-galaxy redshift of $z=1.7825$.
From Image B, we measured an equivalent width of the \ion{Si}{2}\,$\lambda6355$ absorption lines of $56\pm8$\,\AA\ in the rest frame of the SN.
Given the SN Image B phase of \tdbhsiao\ (Section~\ref{sec:sn_phases}), the \ion{Si}{2} velocity and its equivalent width are consistent with those of a normal-velocity SN~Ia or a SN 1991T-like SN~Ia \citep[e.g.,][]{wangwangfilippenko2013,zhaomaedawang2021}.


\begin{table}
\centering
\caption{\textbf{SN templates that fit the observed spectra.}} 
\begin{tabular}{lccccc}
\hline\hline
\textbf{Template} & \textbf{Type} & \textbf{Subtype} & \textbf{Phase} & \textbf{$A_V$} & $\chi^2_\nu$ \\
\hline
\textbf{Image B} & & & & & \\
SN 2013dy & Ia & Ia-norm & $\tbsfa$ & 0.4 & 1.71 \\
SN 2012cg & Ia & Ia-norm & $\tbsfb$ & 0.0 & 2.18 \\
SN 2000cx & Ia & Ia-pec & $\tbsfc$ & 0.0 & 2.60 \\
SN 2013dy & Ia & Ia-norm & $\tbsfd$ & 0.0 & 2.70 \\
SN 2012fr & Ia & Ia-norm & $\tbsfe$ & 1.8 & 2.77 \\
\hline
 \textbf{Image C} & & & & & \\
SN 1994D & Ia & Ia-norm & $\tcsfa$ & 0.1 & 0.69 \\
SN 2013dy & Ia & Ia-norm & $\tcsfb$ & 1.3 & 0.76 \\
SN 2012cg & Ia & Ia-norm & $\tcsfc$ & 0.8 & 0.77 \\
SN 1991T & Ia & Ia 91T-like & $\tcsfe$ & 0.6 & 0.78\\
SN 2011by & Ia & Ia-norm & $\tcsfd$ & 0.3 & 0.79 \\
\hline
\end{tabular}
\raggedright
\tablecomments{Columns show template names, types, subtypes, phases (rest-frame days after peak brightness) for image B or C as indicated, host extinction $A_V$, and reduced chi-square of the fit to the SN H0pe spectrum. Rows show the five best-match SN templates from the {\tt NGSF} analysis.}
\label{tab:superfit}
\end{table}

\section{Methods of measuring phases of the SN images}
\label{sec:sn_phases}

We employ \spw{three} methods to constrain the phases (defined as rest-frame days after peak $B$-band brightness) 
of the SN images. The first method is fitting the Hsiao07 \citep{hsiao2007} spectral templates to the NIRSpec data using Markov-chain Monte Carlo (MCMC) sampling.
Hsiao’s \spw{templates} were constructed from a library of $\sim$600 spectra of $\sim$100 SNe Ia.  \spw{The second method is similar except that the fitting instead uses the SALT3-NIR \citep{salt3-nir} spectral SN Ia models.}  
The original SALT3 model \citep{salt3} was developed from $\sim$1200 spectra of 380 distinct SNe~Ia. To create the SALT3-NIR model, an additional 166 SNe Ia with near-infrared data were incorporated, extending the model's reach to 2\,\micron. \spw{The third method} is to apply the Supernova Identification ({\tt SNID}) software package \citep{snid}, which cross-correlates an observed spectrum with template SN spectra. The {\tt SNID} approach, in contrast to the others,
removes the continuum, and the correlation reflects the features of the spectrum as opposed to the overall shape of the spectral energy distribution. 
The details of the three analyses are described below, and
Appendix~\ref{appendix:sn_phase_results} presents the constraints on the phases of the SN images from these analyses, before accounting for systematic uncertainties described in Section~\ref{sec:method_validation}.

All of constraints on the phases were obtained while the authors were blind to those obtained from fitting the images' light curves (Pierel et al.~2024, submitted). Therefore the analysis presented herein is entirely independent of the light curve phases. 
\spw{Only after completion of both analyses were the two}
sets of results
unblinded by a third party and an inference for $H_0$ made (Pierel et al.~2024, submitted).

\subsection{Hsiao07 SN Ia spectral template}
\label{sec:hsiao_fitting}
\begin{figure*}
\centering
\includegraphics[angle=0,width=6.8in]{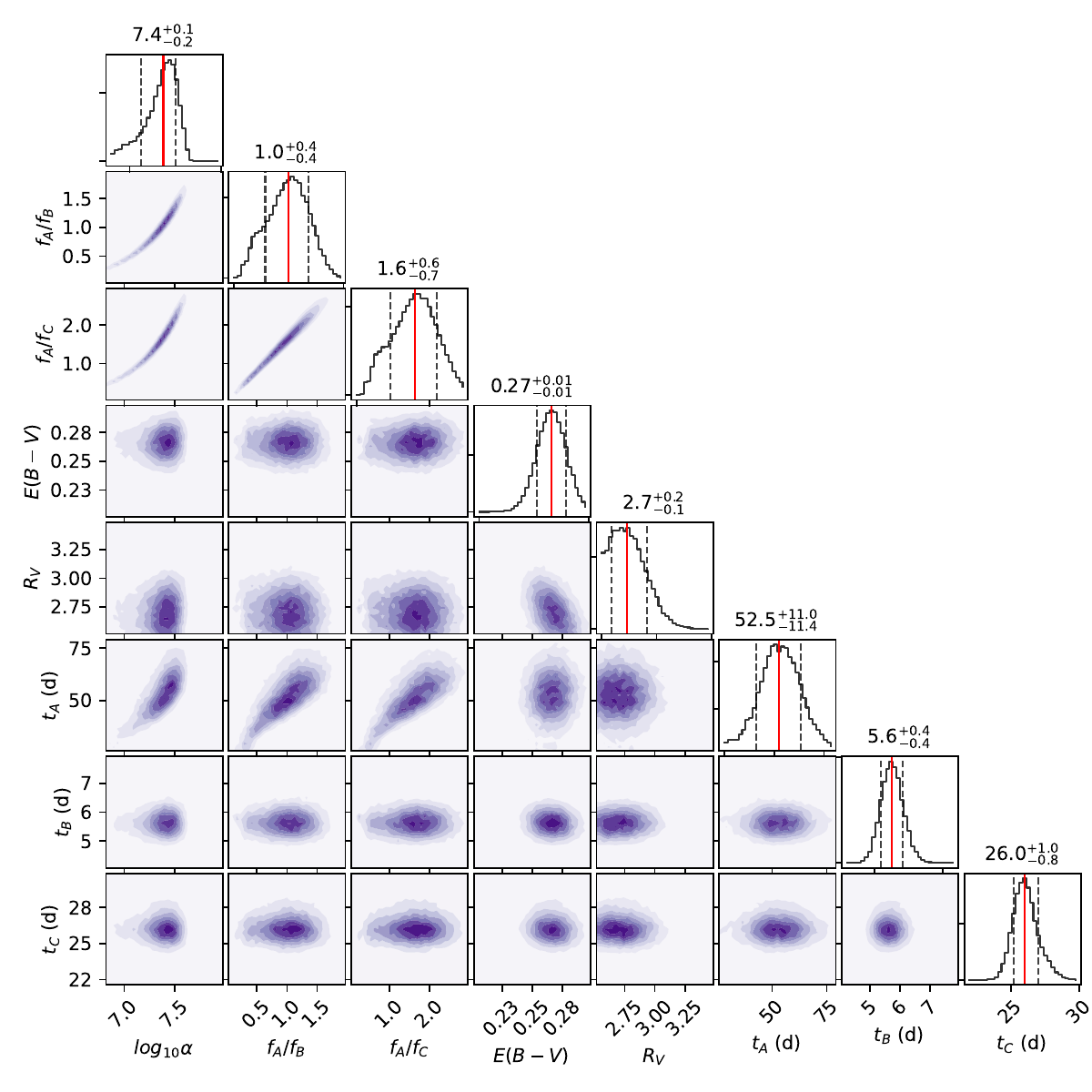}
\caption{Parameter distributions from the Hsiao07 template fits.  $\alpha$, $R_V$, and $E(B-V)$ are model parameters (Section~\ref{sec:hsiao_fitting}), $f_A/f_B$ and $f_A/f_C$ are template-flux ratios between Images A and B and between Images A and C, respectively, and $t_A$, $t_B$, and $t_C$ are rest-frame phases of Images A, B, and C, respectively. The solid vertical line on each histogram marks the median of the parameter, while the dashed vertical lines denote the 68\% confidence interval of each distribution.}
\label{fig:cornerplot_hsiao}
\end{figure*}

\begin{figure*}
\centering
\includegraphics[angle=0,width=5.5in]{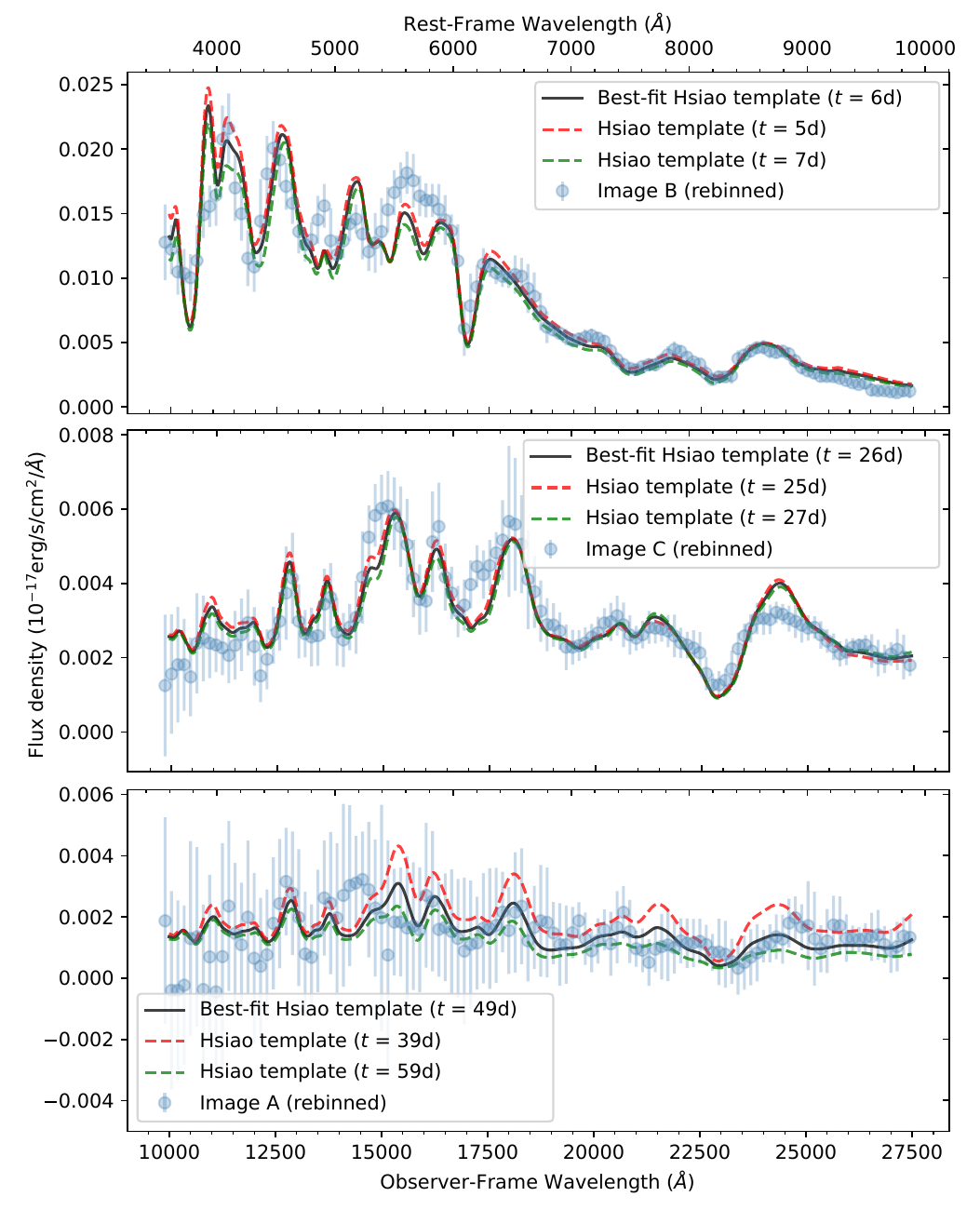}
\caption{Best-fit Hsiao07 templates from Figure~\ref{fig:snspec} are depicted again here, including the models for SN phases that deviate by $\pm 1\sigma$ (rounded to the nearest integers) from the best-fit phases. Data points and error bars show the observed (rebinned  to 120\,\AA) SN~H0pe spectra.}
\label{fig:hsiao_fitting_1sigma_mcmc}
\end{figure*}

We fit the spectra of the three SN images simultaneously with a shared set of parameters. The fit includes a total of eleven free parameters: the overall normalization $\alpha$; the dust-extinction parameters $E(B-V)$ and $R_V$; the pair of flux ratios between \brenda{Images A and B ($f_A/f_B$) and between Images A and C ($f_A/f_C$)}; the image phases $t_A$, $t_B$, and $t_C$ at the time of observation; and a free background value for each of the three spectra. We use the {\tt emcee} software package \citep{emcee} to sample the model parameters and the implementation of the Hsiao07 template in the {\tt SNCosmo} package \citep{sncosmo}.
As a test, we also fit the Hsiao07 template to each SN spectrum separately. Each of these fits had four model parameters: $\alpha$, $E(B-V)$, $R_V$, and $t_i$, where $t_i$ is the phase of that image. 
These results are also listed in Table~\ref{tab:td_results} and are consistent with the shared-parameter results.
Table~\ref{tab:td_results} gives the derived phases, and the black curves in Figure~\ref{fig:snspec} show the best fits. The posterior distributions of the fitted parameters are shown in Figure~\ref{fig:cornerplot_hsiao}, and Figure~\ref{fig:hsiao_fitting_1sigma_mcmc} \spw{illustrates the fit uncertainties}.

In the SN spectrum of Image B, as shown in Figures~\ref{fig:snspec} and \ref{fig:hsiao_fitting_1sigma_mcmc}, the most notable deviation between the observed data and the best-fit model occurs in the rest-frame wavelength range of 5\,500--6\,000\,\AA. This discrepancy could be attributed to the diversity in the continuum level of SNe Ia near the \ion{Si}{2}\,$\lambda5972$ feature or potentially to incomplete modeling of magnesium and sodium features during host-galaxy subtraction. The latter might suggest a color gradient within the host galaxy of SN H0pe. Given that our current extraction technique does not account for the color gradient of extended sources, we anticipate that future spectroscopic observations of the host galaxy, conducted in the absence of the SN, will provide insights into this observed discrepancy.

\subsection{SALT3-NIR model}
\label{sec:salt_fitting}

For the SALT3-NIR \citep{salt3-nir} analysis, the input parameters for this model are the overall normalization $x_0$, a light-curve shape parameter (``stretch'') $x_1$, a color parameter $c$, and the phase and background parameters as for the analysis using the Hsiao07 template\null. Because the SALT3-NIR model \spw{is limited to phases $\le$50 days,}
\spw{for larger phases,} we assigned likelihoods  equal to the 50-day value.
\spw{As for the Hsiao07 templates, we tested independent fits to each spectrum in addition to the primary simultaneous fit to all three spectra.}
Figure~\ref{fig:snspec} shows the best fits of the SALT3-NIR model to the spectra, \spw{and Table~\ref{tab:td_results} lists the best-fit image phases}.
Appendix~\ref{appendix:salt3nir_results} presents the full results of the SALT3-NIR model-fitting \spw{including parameter distributions and uncertainties.} 

The SALT3-NIR results are statistically consistent with those from the Hsiao07-template analysis. However, the phase of Image A from the SALT3-NIR fitting approaches the 50-day upper \spw{validity} limit of the model. For the joint fitting, this proximity to the limit of the model may affect the SALT3-NIR model's constraints on the other image phases, potentially leading to biased estimates of the time delays and magnifications. Therefore, we do not rely on the joint SALT3-NIR fitting for our primary constraints on the phases or relative time delays of the images.

\subsection{{\tt SNID}}

\begin{figure*}
\centering
\includegraphics[angle=0,width=6in]{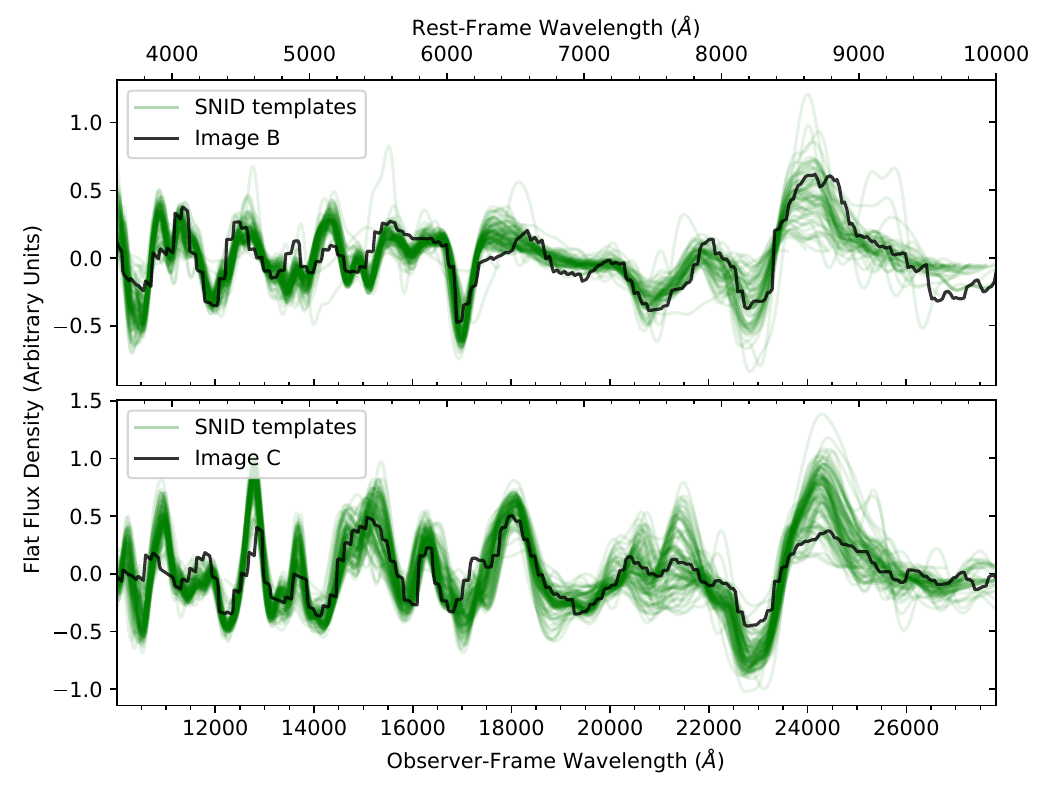}
\caption{Flattened flux density of SN Images B and C for the {\tt SNID} analysis \brenda{(black)}. Green curves show the 100 best-fitting {\tt SNID} templates.}
\label{fig:snid}
\end{figure*}

As a third measurement, we derive constraints {\em independent} of the SN color, i.e., using only the spectral features. For this, we employed {\tt SNID} \citep{snid}, which removes the continuum from the SN spectrum before cross-correlating the SN spectrum with template spectra. We used both the built-in {\tt SNID} template library (``template 2.0'') and a library that we constructed from the SALT3-NIR SN Ia template,
which includes 1\,750 spectra across 70 SN phases, each spanning 25 different combinations of the stretch ($x_1$) and color ($c$) parameters.
%
%
Figure~\ref{fig:snid} plots the continuum-removed spectra of Images B and C superimposed on the 100 best-fitting SN spectra. 

With {\tt SNID}'s default template-matching thresholds, 176 templates---all SN Ia---match the Image B spectrum. The \spw{matching templates all have early phases (Table~\ref{tab:td_results}) but are consistent with the Hsiao07 and SALT3-NIR results within the uncertainties (Table~\ref{tab:td_results})}. 
There are 163 \spw{matching} templates that are in the Ia-norm subtype, \spw{all giving essentially the same phases but  uncertainties larger than those from Hsiao07 or SALT3-NIR.}
The Image C \spw{spectrum} has 165 matching templates,  164 of them SNe Ia, and
107 with Ia-norm templates. 
No \spw{phase or even a} favored SN type can be identified from the low signal-to-noise spectrum of Image A\null. 
\spw{Using the simulated templates generated from the SALT3-NIR model
gives phases and uncertainties similar to those from the templates 2.0 package but with a preference for a slightly later phase for Image~B (Table~\ref{tab:td_results}).}

The {\tt SNID} analysis provides decisive evidence that SN H0pe is Type Ia. All SN phases from the {\tt SNID} analysis are consistent with the values from the SALT3-NIR and Hsiao07 analyses.
%

\section{Uncertainties of the measurement of time delays and magnifications}
\label{sec:method_validation}

Image A might be at a phase outside the SALT3-NIR model's valid range, and the image is too faint for SNID to reliably infer its phase. This leaves the Hsiao07 templates as the principal basis for our analysis. To assess the templates' ability to determine SN phases,
our analysis began with joint fitting of a set of well-observed, nearby SNe Ia with the Hsiao07 templates, as described in Section~\ref{sec:simulation}.
Section~\ref{sec:lensing} presents the subsequent modeling of the effects of millilensing and microlensing to assess those processes' impacts on the time delay and magnification measurements.
Final constraints on the SN phases, magnifications, and relative time delays are presented in Section~\ref{sec:final_constraints}.

\subsection{Simulation of the Hsiao07 modeling}
\label{sec:simulation}

To assess the ability of the Hsiao07 template to recover the phase of the SN images, we applied the same analysis to the spectra of a set of well-observed, nearby SNe Ia listed in  {\tt SNID}’s built-in library. 
%
\begin{figure*}
\centering
\includegraphics[angle=0,width=6.2in]{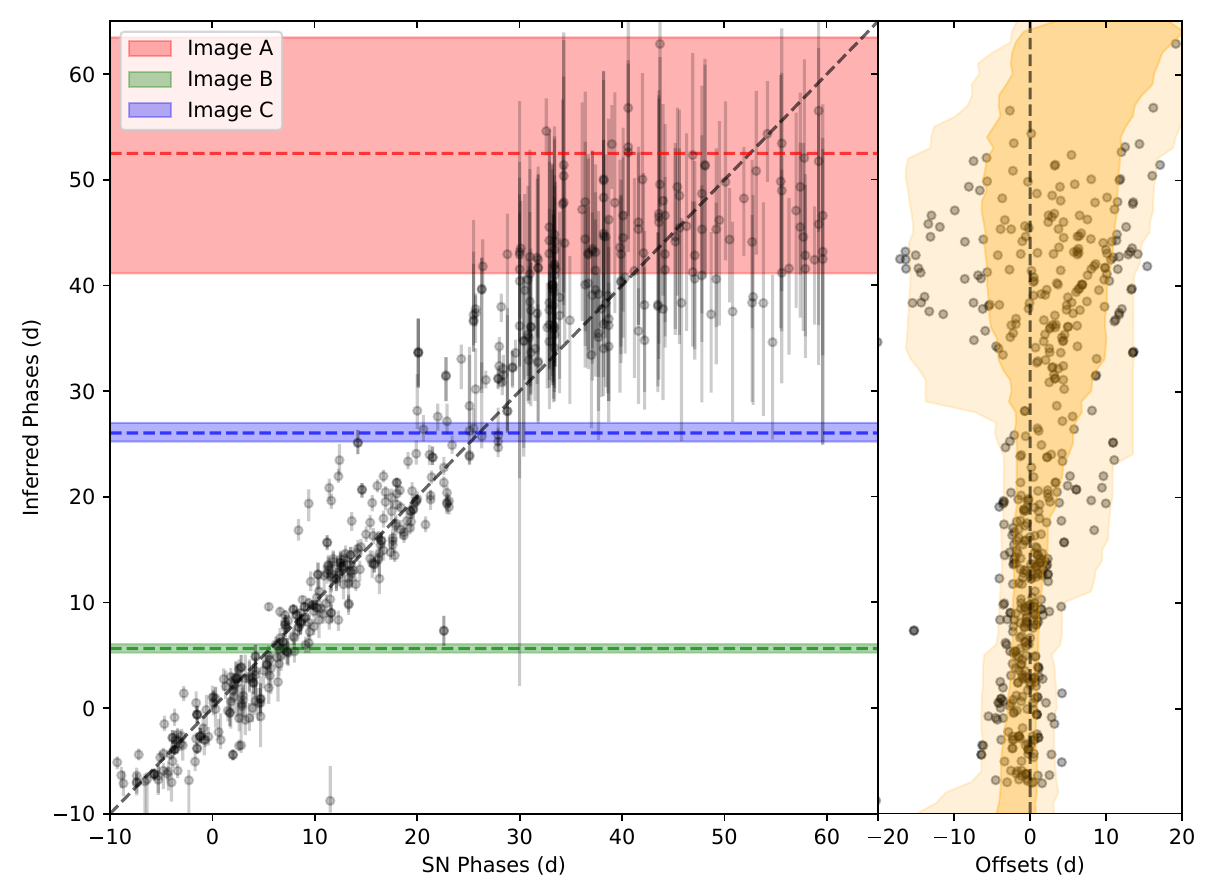}
\caption{{\bf Left:} inferred phases from Hsiao07 fitting versus actual phases of a set of nearby SNe Ia. Error bars correspond to the 68\% confidence intervals of the inferred phases. Horizontal dashed lines and shaded regions mark the inferred phases and their 68\% confidence intervals for the three images of SN H0pe. The diagonal dashed line shows equality. {\bf Right:} deviation of the best-fit inferred phase from the actual phase. Shaded regions mark the 68\% (darker) and 95\% (lighter) confidence intervals of the residuals measured within a $10$-day rolling window.  Simulations including input SNe beyond 60 days could, in principle, increase the uncertainty for Image A, and constraints will benefit from future simultaneous measurement from spectra and imaging. }
\label{fig:validation_hsiao}
\end{figure*}
%
The SNe were chosen to have at least one spectrum in each of the following ranges of phase relative to maximum: ($-10$, 10), (10, 30), and (30, 60) days. These ranges correspond to the values inferred for SN H0pe Images B, C, and A, respectively, using the Hsiao07 template, which includes spectra through 80 days.
This gave 228 sets of spectra for the simulation with each set consisting of three SN spectra in the three phase bins.
To each spectrum, we added the residuals of the NIRSpec spectra from the smoothed spectra (as shown by the dash-dotted lines in Figure~\ref{fig:snspec}), drawing randomly from the residual distribution across all wavelengths. This step \spw{assumes the residuals are random rather than correlated with wavelength.}
Additionally, we rescaled the spectra to match the signal-to-noise ratio of the NIRSpec observations.

Figure~\ref{fig:validation_hsiao} compares the actual and inferred phases.  
The residuals and the confidence intervals, shown by shaded areas, are plotted in the right panels of Figure~\ref{fig:validation_hsiao}. As shown in Figure~\ref{fig:validation_hsiao}, the inferred phase is generally in agreement with the true value, and the agreement is especially good at phases $<$30~days. Beyond 30 days, the uncertainty associated with the inferred phase  is significantly greater, primarily because of Image A's low signal-to-noise ratio and the slow evolution of SN spectral features at these phaess. 



\subsection{Millilensing and microlensing}
\label{sec:lensing}

Lens-model predictions of the image magnification ratios and relative time delays include only structure on the scale of galaxy members and the cluster itself. Therefore, the effects of millilensing by subhalos and microlensing by stars and compact objects have to be included in the uncertainties on the inferred phases and time delays.
%
Millilenses  affect only the magnifications of the SN images because the SN photosphere (${\sim}10^{14}$ to $10^{15}$ cm in radius) is much smaller than the scale of a millilens caustic. Microlenses, such as stars and stellar remnants, on the other hand, have Einstein radii comparable to the size of the SN photosphere. As the expanding SN photosphere crosses microlens caustics, different portions of the photosphere can be magnified differently, leading to wavelength-dependent changes in the SN spectrum \citep{goldsteinnugentkasen2018,foxleymarrablecollettvernardos2018,hubersuyunoebauer2019}. Consequently, microlensing can affect not only the magnification of the SN images but also time-delay estimates.


To compute the uncertainties due to microlensing and millilensing, we utilized 4\,000 simulations of SNe Ia spectra. (The simulations are described in detail in the companion paper presenting time-delay inferences from the photometry of the SN: Pierel et al.~2024, submitted.) In brief, for each of the four theoretical models of microlensing caustics from 
\citet{suyuhubercanameras2020} and \citet{hubersuyunoebauer2021}, a set of SNe Ia were placed at 1\,000 random positions in the source plane.
The simulated spectrum at each position compared to the input spectrum without microlensing was used to determine the wavelength-dependent magnification due to microlenging. 
The millilens magnification distributions were calculated for a range of dark-matter subhalo mass functions and substructure fractions anticipated from theory as well as observations \citep{gilmanbirrernierenberg2020}. 
%
A magnification drawn from one of the millilensing distributions was applied to each simulated SN sight line. As in Section~\ref{sec:simulation}, we added noise to the simulated spectra to match the signal-to-noise ratios of the three SN H0pe spectra. Then we simultaneously fit the set of three simulated spectra corresponding to the three SN H0pe images. The differences between the inferred and input SN phases and magnifications yielded an estimate of the systematic uncertainties arising from millilensing and microlensing. Table~\ref{tab:lensing_effects}
gives the results.

\begin{table}
	\centering
	\caption{{Microlensing and Millilensing  Uncertainties}}
	\begin{tabular}{cccc}
		\hline
$\sigma_{m}$ ($\pm$ AB magnitude) & Image A & Image B & Image C \\ 
		\hline
Microlensing & 0.04 & 0.12 & 0.04 \\
Millilensing & 0.04 & 0.04 & 0.04 \\ 
        \hline
	\end{tabular}
\raggedright
\tablecomments{Median standard deviation of magnitude changes $\sigma_{m}$  across the rest-frame wavelength range 3\,600\,\AA\ to 10\,000\,\AA\ in the simulated SN Ia sample.}
	\label{tab:lensing_effects}
\end{table}

\section{Constraints on time delays and magnifications}
\label{sec:final_constraints}



For our joint fitting using the Hsiao07 template, for each $i$-th step in the MCMC chain from our joint fitting, we obtained a set of eight model parameters, denoted as $\mathbf{X}^i=[\alpha_{A}^i, \alpha_{B}^i, \alpha_{C}^i, E(B-V)^i, R_V^i, t_A^i, t_B^i, t_C^i]$. The alphas are related to the ones described in Section~\ref{sec:hsiao_fitting} by $\alpha_{A}^i=\alpha^i$, $\alpha_{B}^i=\alpha^i/(f_A/f_B)^i$, $\alpha_{C}^i=\alpha^i/(f_A/f_C)^i$. 
We proceeded by randomly selecting an SN among the SNe Ia with inferred phase within a $\pm$10-day window and calculating the median offsets of this SN's inferred phase from its actual value. Next, we chose a set of three SN spectra from our simulated millilensing and microlensing samples and determined the median offsets  between the input and inferred parameter values. These offsets, derived from the simulations, were then applied to the parameters in the MCMC chain. This process resulted in a modified set of eight parameters  $\mathbf{Y}^i$, where $Y_1^i=\alpha_{A}^i+\delta \alpha_{A}^j+\delta \alpha_{A}^k$, $Y_2^i=\alpha_{B}^i+\delta \alpha_{B}^j+\delta \alpha_{B}^k$, $Y_3^i=\alpha_{C}^i+\delta \alpha_{C}^j+\delta \alpha_{C}^k$, $Y_4^i=E(B-V)^i+\delta E(B-V)^k$, $Y_5^i=R_V^i+\delta R_V^k$, $Y_6^i=t_A^i+\delta t_A^j+\delta t_A^k$, $Y_7^i=t_B^i+\delta t_B^j+\delta t_B^k$ and $Y_8^i=t_C^i+\delta t_C^j+\delta t_C^k$. Here the $j$ and $k$ \brenda{indices are} for the set of randomly selected SN spectra from the low-redshift SN Ia simulation and from the millilensing and microlensing simulation, respectively. For the simulation of the low-redshift SNe Ia, we assume that the extinction parameters $R_V$ and $E(B-V)$ were fit accurately for each SN. 


\begin{figure*}
	\centering
	\includegraphics[angle=0,width=6.5in]{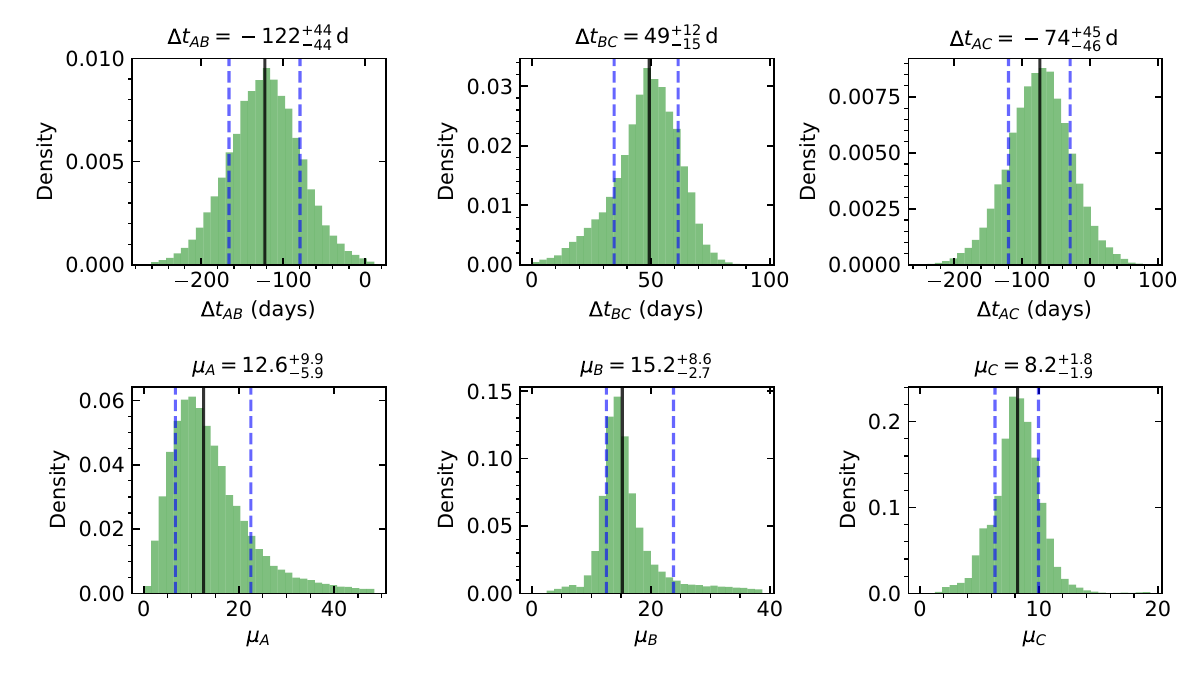}
	\caption{Posterior probability densities of the relative time delays and the macro lens magnifications between SN H0pe's images. Histograms show values derived from the Hsiao07 templates. The distributions integrate the uncertainty from the MCMC fitting with systematic uncertainties, estimated from the model fitting of nearby SN Ia spectra and from the simulated SN Ia spectra. The uncertainties take into account the impacts of microlensing and millilensing effects.}
	\label{fig:pstr_hsiao}
\end{figure*}


To constrain the absolute magnifications of the three images of the SN, we used the MCMC sample from our joint fitting. For each set of modified model parameters $\mathbf{Y}^i$ in the MCMC chain at the $i$-th step, we generated a set of three synthetic SN spectra using the {\tt SNCosmo} package \citep{sncosmo}.

The expected rest-frame $B$-band magnitudes of SNe Ia were obtained from the complete sample of \citet{scolnic2018}. To estimate $\hat{m}_B$ at $z=1.782$ independent of the cosmological distance and $H_0$, we used a polynomial fit to interpolate $\hat{m}_B$ as a function of the logarithmic redshift, which yields $\hat{m}_B=26.32\pm0.03$ at $z=1.782$.
As a cross-check performed after unblinding, a concordance cosmology described by the $\Lambda$ cold dark matter model ($\Lambda$CDM) with $\Omega_m=0.3$, $\Omega_\Lambda=0.7$, and a Hubble constant $H_0=70$~\kms~{Mpc}$^{-1}$ yields $\hat{m}_B=26.30$ at the redshift.

The magnitude in a specific band for a given amplitude of the model is dependent upon host-galaxy extinction. In our analysis using the Hsiao07 templates, we assumed that SN H0pe is a standard candle in the rest-frame $Y$-band, where previous studies \citep[e.g.,][]{avelino_type_2019} show that the SN brightness is not substantially correlated with the spectral stretch and color and is homogeneous without these corrections. 
We then determined the expected $Y$-band magnitude for an unmagnified SN Ia. For an SN Ia spectrum produced by the Hsiao07 template with the best-fit dust-extinction parameters, we varied the normalization factor of the spectrum to match its $Y$-band magnitude to that of an SN Ia without extinction and with a rest-frame $B$-band magnitude ($\hat{m}_B$) equal to the expected value. 
This yielded an unlensed $Y$-band magnitude of $27.15\pm0.03$.
Subsequently, we synthesized a $Y$-band measurement of the SN for each set of parameters $\mathbf{Y}^i$ in the MCMC chain and calculated the predicted $Y$-band magnitude at the SN phase $t=0$ for each image based on the Hsiao07 template. We then determined the magnification by comparing the resultant $Y$-band magnitude to the expected value.
Table~\ref{tab:y_mag} list the resultant $Y$-band magnitudes derived from the MCMC chain and the predicted magnitudes at $t=0$.

\begin{table}
	\centering
	\caption{\textbf{$Y$-band apparent AB magnitudes of the SN images.} }
	\begin{tabular}{crrr}
		\hline\hline
		$i=$&		A & B & C \\ 
		\hline
		\multicolumn{4}{c}{Magnitudes from Hsiao07 fits}\\
		& $25.41^{+0.45}_{-0.32}$ & $24.60^{+0.05}_{-0.05}$ & $24.84^{+0.06}_{-0.06}$ \\
		\multicolumn{4}{c}{Predicted magnitudes for $t_i=0$}\\
		& $24.39^{+0.69}_{-0.63}$ & $24.18^{+0.22}_{-0.48}$ & $24.85^{+0.29}_{-0.21}$ \\
        \hline
	\end{tabular}
	\raggedright
	\tablecomments{The rest-frame $Y$-band photometry was synthesized using the Dark Energy Camera's $y$ filter \citep{2018ApJS..239...18A} set at airmass 1.3 based on the Hsiao07-template fits. Phases $t_i$ are rest-frame days after the peak brightness in the rest-frame $B$-band.}
	\label{tab:y_mag}
\end{table}

Finally, we reconstructed the posterior distribution of the time delays and the macro lens magnifications (i.e., the magnifications from the cluster lens) as illustrated in Figure~\ref{fig:pstr_hsiao}. Table~\ref{tab:correlation_hsiao} lists the correlation matrices of the relative time delays, magnification ratios, and magnifications from our analysis using the Hsiao07 template. 
Our final constraints on the SN phases, relative time delays, and magnifications are listed in Table~\ref{tab:final_result}.

\begin{deluxetable*}{lRRRRRRRRR}
	\centering
	\tablecaption{\textbf{Correlation matrices of model parameters}}
\tablehead{&
\colhead{$\Delta t_{\rm AB}$}& 
\colhead{$\Delta t_{\rm BC}$}& 
\colhead{$\Delta t_{\rm AC}$}& 
\colhead{$\mu_A/\mu_B$}& 
\colhead{$\mu_B/\mu_C$}& 
\colhead{$\mu_A/\mu_C$}& 
\colhead{$\mu_A$}& 
\colhead{$\mu_B$}& 
\colhead{$\mu_C$}}
\startdata
\multicolumn{10}{l}{Hsiao07}\\
{$\Delta t_{\rm AB}$}&1.000  \\
{$\Delta t_{\rm BC}$}&-0.002 & 1.000  \\
{$\Delta t_{\rm BC}$}& 0.947 & 0.320 & 1.000 \\
{$\mu_A/\mu_B$}&-0.572 & 0.086 & -0.514 & 1.000\\
{$\mu_B/\mu_C$}&-0.268 & -0.457 & -0.402 & -0.178 & 1.000 \\
{$\mu_A/\mu_C$}&-0.633 & -0.304 & -0.698 & 0.420 & 0.678 & 1.000 \\
{$\mu_A$}&-0.721 & -0.131 & -0.725 & 0.480 & 0.525 & 0.872 & 1.000 \\ 
{$\mu_B$}&-0.322 & -0.223 & -0.394 & -0.202 & 0.851 & 0.535 & 0.622 & 1.000  \\
{$\mu_C$}&-0.020 & 0.651 & 0.191 & 0.009 & -0.800 & -0.321 & 0.021 & 0.022 & 1.000\\
\tableline
\multicolumn{10}{l}{SALT3-NIR}\\
{$\Delta t_{\rm AB}$}&1.000   \\
{$\Delta t_{\rm BC}$}&-0.082 & 1.000  \\
{$\Delta t_{\rm BC}$}&0.932 & 0.284 & 1.000  \\
{$\mu_A/\mu_B$}&-0.467 & 0.065 & -0.426 & 1.000 \\
{$\mu_B/\mu_C$}&-0.055 & -0.243 & -0.141 & -0.409 & 1.000\\
{$\mu_A/\mu_C$}&-0.516 & -0.244 & -0.585 & 0.516 & 0.317 & 1.000 \\
{$\mu_A$}&-0.639 & -0.030 & -0.626 & 0.683 & 0.101 & 0.797 & 1.000  \\ 
{$\mu_B$}&-0.053 & -0.120 & -0.095 & -0.477 & 0.898 & 0.127 & 0.112 & 1.000  \\
{$\mu_C$}&0.018 & 0.588 & 0.231 & 0.014 & -0.357 & -0.458 & 0.001 & -0.070 & 1.000
\enddata
\label{tab:correlation_hsiao}
\tablecomments{Each $i$-th row and $j$-th column is the correlation between $i$-th and $j$-th parameters. 
The upper section of the table is for the Hsiao07 models (Section~\ref{sec:hsiao_fitting}), and the lower section is for the SALT3-NIR models (Section~\ref{sec:salt_fitting}).
Parameters are the relative time delays $\Delta t_{AB}$, $\Delta t_{BC}$, $\Delta t_{AC}$, the magnification ratios $\mu_A/\mu_B$, $\mu_B/\mu_C$, $\mu_A/\mu_C$, and the magnifications $\mu_A$, $\mu_B$, $\mu_C$. }
\end{deluxetable*}

\begin{table}
	\centering
	\caption{\textbf{Summary of final constraints.} }
	\begin{tabular}{cccc}
		\hline\hline
$i=$&		A & B & C \\ 
		\hline
\multicolumn{4}{l}{Hsiao07 template fits}\\
$t_i$	& \tdahsiao & \tdbhsiao & \tdchsiao \\
$t_i$	& $54.5_{-17.0}^{+18.7}$ d & $6.2_{-1.7}^{+2.5}$ d & $24.5_{-3.4}^{+3.6}$ d \\
$\mu_i$&   \muahsiao & \mubhsiao & \muchsiao \\ 
$\mu_i$&  $14.5_{-7.1}^{+13.3}$ & $15.3_{-2.7}^{+8.2}$ & $8.4_{-1.7}^{+1.7}$ \\ 
\multicolumn{4}{l}{SALT3-NIR fits}\\
$t_i$	& \tdasalt & \tdbsalt & \tdcsalt \\
$\mu_i$&   \muasalt & \mubsalt & \mucsalt \\ 
$\mu_i$&   \muaysalt & \mubysalt & \mucysalt \\ 
\hline
$i=$& AB & BC&AC\\
\hline
\multicolumn{4}{l}{Hsiao07 template fits}\\
$\Delta t_i$ &  \tdabhsiao & \tdbchsiao & \tdachsiao \\
$\Delta t_i$ &  $-133.8_{-51.1}^{+47.2}$ d & $50.3_{-12.7}^{+12.0}$ d & $-82.8_{-54.6}^{+50.2}$ d \\
\multicolumn{4}{l}{SALT3-NIR fits}\\
$\Delta t_i$ &  \tdabsalt & \tdbcsalt & \tdacsalt \\
\hline
	\end{tabular}
\raggedright
\tablecomments{Phases $t_i$ are rest-frame days after the peak brightness. 
Magnifications $\mu_i$ are the macro-magnifications from the cluster and are based on SN Ia absolute magnitudes as described in Section~\ref{sec:final_constraints} for the Hsiao07 templates and Section~\ref{sec:constraints_salt} for the SALT3-NIR fits. For the Hsiao07 template fits, the second row of $t_i$, $\mu_i$, and $\Delta t_i$ are based on a post-unblinding test where we expand the third phase bin for selecting the nearby SN Ia spectra from (30, 60) days to (30, 80) days (as described in Section~\ref{sec:hsiao80}).
For the SALT3-NIR fits, the first row of $\mu_i$ is based on $B$-band absolute magnitudes, and the second row is based on $Y$-band.
Relative time delays $\Delta t_i$ are days in the observer frame.
}
	\label{tab:final_result}
\end{table}

\section{A post-unblinding test}
\label{sec:hsiao80}

\begin{figure*}
	\centering
	\includegraphics[angle=0,width=5.5in]{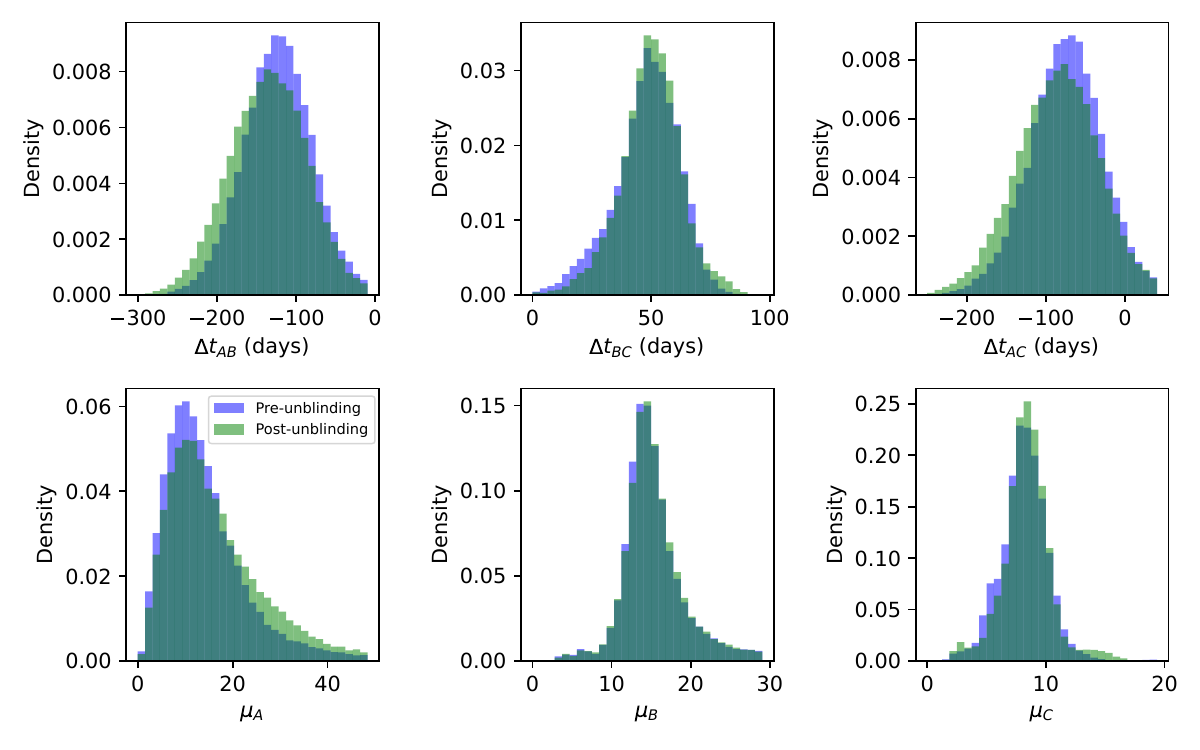}
	\caption{Posterior probability densities of the relative time delays and the macro lens' magnifications between SN H0pe's images. For the pre-unblinding result, we simulated the joint Hsiao07-template fitting by selecting a set of three spectra from three SN phase bins: ($-10$, 10), (10, 30), and (30, 60) days (as described in Section~\ref{sec:simulation}). For the post-unblinding test, we expand the third phase bin to (30, 80) days.}
	\label{fig:hsiao_60vs80}
\end{figure*}

As described in Section~\ref{sec:simulation}, we chose three bins in SN phase to select the spectra of nearby SNe Ia. These bins were determined {\it a priori} based on the result of the joint MCMC fitting, before we conducted the simulation. Given the low signal-to-noise ratio of Image A's spectrum, there is a lack of sensitivity to phases for SN Ia spectra with inferred phases beyond $\sim40$ days. To assess the impact of the upper limit of the third phase bin (assigned for Image A) on the inferred time delays and magnifications, we expanded the range of third phase bin from (30, 60) days to (30, 80) days. We then performed the same simulation as described in Section~\ref{sec:simulation} with the extended phase bin, while keeping the results of other simulations unchanged to determine the final constraints. Table~\ref{tab:final_result} lists the resulting constraints on the phases, the relative time delays, and the magnifications of SN H0pe's images. Figure~\ref{fig:hsiao_60vs80} shows the distributions of these parameters, over-plotted on the pre-unblinding distributions.
This test was conducted after the unblinding of the result from the $H_0$-inference team (Pascale et al.~2024, in prep.). 

As shown in Figure~\ref{fig:hsiao_60vs80}, the distributions of relative time delays and magnifications remain largely unaffected by the adjustment of the upper limit, for which the deviations of the corresponding distributions are much smaller compared to the uncertainties of these measurements.
Given the slightly longer time delays compared to the original result, the inferred $H_0$ will be $\sim3$\% smaller than that based on the pre-unblinding constraints. However, the difference is only $\sim0.1\sigma$, which is not statistically substantial. For the $H_0$ inference by Pascale et al.~2024 based on the time-delay measurements from both the spectroscopic and photometric analyses, the impact of the upper limit of the phase bin for our simulation is minimal ($<1$\%).  


\section{Conclusions}
\label{sec:discussion}

All of our analyses of the spectra of SN H0pe classify SN H0pe as a Type Ia SN and favor a ``normal'' \citep{snid} subtype. The velocity of the \ion{Si}{2}\,$\lambda6355$ absorption line is consistent with a normal-velocity SN Ia. 
\spw{The SN phases and relative time delays are constrained with 1-$\sigma$ uncertainties of 2-4 days for Images B and C and two weeks for the latest-arriving Image A.} 

Our analyses using the Hsiao07 template, the SALT3-NIR model, and the {\tt SNID} software package yield consistent results for the phases of the SN images. The relative time delays between the three SN H0pe images and their magnifications, as derived from our analyses based on the Hsiao07 template and the SALT3-NIR model, are statistically in agreement. However, the inferred phase of Image A is near or above the upper boundary of the SALT3-NIR model's valid range. During the joint fitting, this affected the constraints of the SALT3-NIR model, potentially biasing the estimates of time delays and magnifications.
The SNID model approach, which was unable to utilize the information from the faintest Image A, \spw{yielded larger uncertainties than} the Hsiao07 analysis but only uses the spectral features, since the SN continuum is subtracted.

In addition to time delays and magnifications, the fits using the Hsiao07 template constrain the host extinction $E(B-V)=0.27\pm0.02$ and $R_V=2.73\pm0.17$. The $E(B-V)$ is consistent with the values listed in Table~4 of \citet{Frye_2024}. This indicates that the SN exploded in a relatively dusty environment with an extinction $A_V\gtrsim0.7$.


SN H0pe will be valuable not only for measuring the Hubble--Lema\^{\i}tre constant \brenda{but also, because this is the second-highest currently known redshift SN~Ia}, for testing whether SNe at large lookback times had the same properties as SNe today. Based on spectra at multiple rest-frame epochs, SN H0pe matches the most common type of  SN~Ia in the nearby universe.  
Our analysis of the SN images' phases, based solely on the NIRSpec spectra, suggests a relative uncertainty of $\gtrsim 20$\% in the time delay between the two brightest images (B and C). 
This would limit the precision of constraints on  the Hubble constant from the measured spectroscopic time delay to be $\gtrsim 20$\%, given more precise lens model predictions.
Beyond the blind analysis detailed in this paper and the associated photometric-analysis paper (Pierel et al.~2024, submitted), a comprehensive joint analysis---utilizing both  the spectroscopic time delay and  the photometric measurements (Pascale et al.~2024, in prep.)---will yield a combined measurement of the Hubble constant with improved precision.

\clearpage
\begin{acknowledgments}

This work is based on observations made with the NASA/ESA/CSA {\it JWST}. The data were obtained from the Mikulski Archive for Space Telescopes (MAST) at the Space Telescope Science Institute, which is operated by the Association of Universities for Research in Astronomy, Inc., under NASA contract NAS 5-03127 for {\it JWST}. These observations are associated with program {\it JWST} DDT 4446. RAW and SHC acknowledge support from NASA {\it JWST} Interdisciplinary Scientist grants NAG5-12460, NNX14AN10G and 80NSSC18K0200 from GSFC. AZ acknowledges support by Grant No. 2020750 from the United States-Israel Binational Science Foundation (BSF) and Grant No. 2109066 from the United States National Science Foundation (NSF); by the Ministry of Science \& Technology, Israel; and by the Israel Science Foundation Grant No. 864/23. SH thanks the Max Planck Society for support through the Max Planck Research Group for S. H. Suyu and the European Research Council (ERC) under the European Union’s Horizon 2020 research and innovation programme (grant agreement No 771776) for funding. CL acknowledges support from the National Science Foundation Graduate Research Fellowship under grant No. DGE-2233066. P.L.K. acknowledges funding from NSF grants AST-1908823 and AST-2308051.

\end{acknowledgments}

%

\vspace{5mm}
Data used in this paper can be retrieved at MAST: doi: 10.17909/rqdx-3976.

\facilities{{\it JWST} (NIRSpec spectroscopy and NIRCam imaging)}


\software{jwst 1.10.2 (\url{https://jwst-pipeline.readthedocs. io/en/latest/jwst/introduction.html}), MOS Optimal Spectral Extraction (\url{https://spacetelescope.github.io/jdat_notebooks/notebooks/optimal_extraction/Spectral_Extraction-static.html}), sncosmo (\url{https://sncosmo.readthedocs.io/en/stable/}), emcee (\url{https://emcee.readthedocs.io/en/stable/}), NGSF (\url{https://github.com/oyaron/NGSF}), SNID (\url{https://people.lam.fr/blondin.stephane/software/snid/}).}



\bibliography{spec_time_delay}{}
\bibliographystyle{aasjournal}

\clearpage
\appendix
\restartappendixnumbering

\section{Background subtraction, 2D-spectrum modeling, and source extraction}
\label{appendix:2d_spec}

\begin{figure}
	\centering
	\includegraphics[angle=0,width=3.5in]{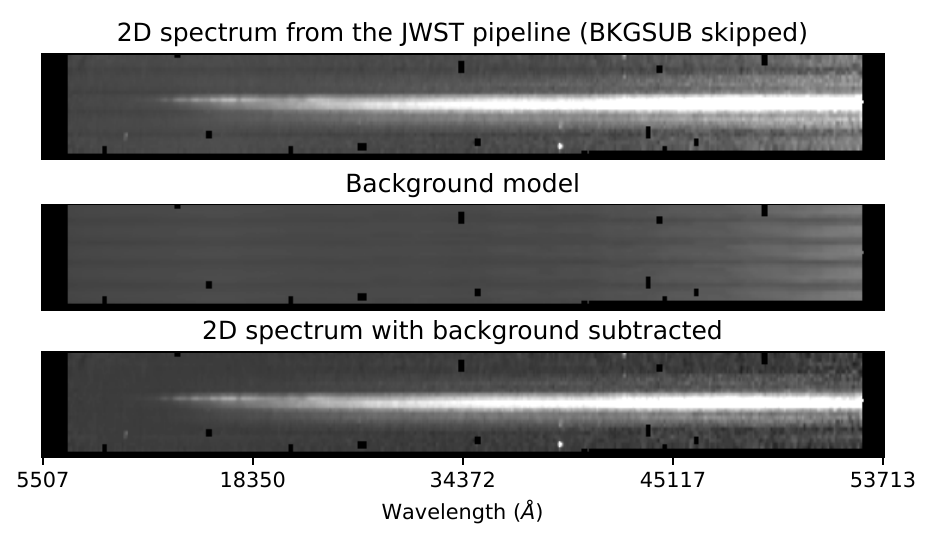}
	\caption{An example of the background subtraction. The top panel shows the 2D spectrum from the {\it JWST} pipeline with the background-subtraction step (BKGSUB) skipped. The middle panel shows the background model from the local-minimum-flux method. The bottom panel shows the residuals after subtracting the background model from the original spectrum.}
	\label{fig:bkgsub}
\end{figure}

Because the MSA slitlets used to observe SN H0pe are occupied by extended sources such as the SN's host galaxy, the default {\it JWST} pixel-to-pixel background subtraction based on in-scene and off-scene nods was skipped. In addition, the lack of ``blank sky'' shutters in this NIRSpec observation makes it hard to obtain a well-defined master background to feed to the {\it JWST} pipeline for subtraction. Instead, we developed a custom code to evaluate a 2D master background using the local minimum flux in 2D spectra for a given wavelength. In detail, for each pixel in a 2D spectrum without a background subtraction, we calculated the minimum flux from the pixels with MSA shutters in the same position along the slit direction. Because the minimum flux could be biased by bad pixels (e.g., cosmic-ray masks), we smoothed the resulting 2D minimum-flux array along the dispersion direction using a median filter with a width of 10 pixels to give a 2D background model. Figure~\ref{fig:bkgsub} shows an example of the background subtraction.

Source extraction was based on the optimal extraction algorithm from \citep{horne1986} and applied to the \spw{background-subtracted} 2D spectra.
For each wavelength in a 2D spectrum, we evaluated the spatial flux distribution by integrating the 2D spectrum \spw{along the slit} within a range of nearby wavelengths. We then fit kernel functions (Moffat functions for extended sources and Gaussian functions for point-like sources) to the flux distribution of the integrated spectra. For extended sources, we obtained a set of best-fit Moffat functions by scanning over the whole wavelength range. Assuming that the core width of the Moffat functions is linear and the power index of the Moffat functions is a constant over the wavelength direction, we constructed 2D extraction models based on the best-fit core widths and power indexes of Moffat functions. For point-like sources, the extraction Gaussian was fitted to the composite PSFs. For unresolved sources from the NIRSpec observation, we assigned a point-source kernel at the source position. Finally, we fixed the shapes of these kernels at each wavelength and fit the kernels flux simultaneously into the original 2D spectra.  

\begin{figure}
	\includegraphics[angle=0,width=3.5in]{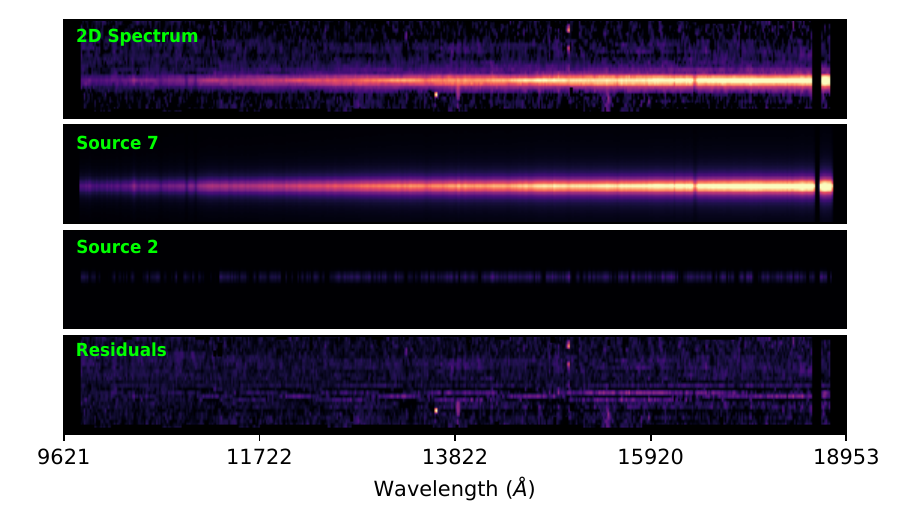}
	\caption{G140M spectrum from Slit A. The top panel shows the 2D spectrum from the {\it JWST} pipeline. The middle two panels show the best-fit models of the host galaxy (Source 7) and the SN (Source 2). The bottom panel shows the residuals after subtracting the galaxy and the SN model from the original spectrum.}
	\label{fig:a_140_2d}
\end{figure}

\begin{figure}
	\includegraphics[angle=0,width=3.5in]{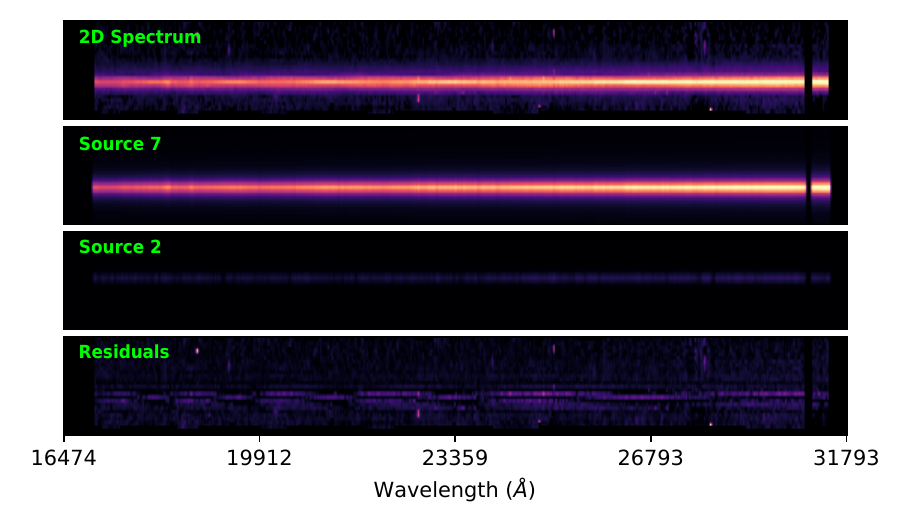}
	\caption{G235M spectrum from Slit A. Top panel is the 2D spectrum from the {\it JWST} pipeline. The middle two panels show the best-fit models of the host galaxy (Source 7) and the SN (Source 2). The bottom panel shows the residuals after subtracting the galaxy and the SN models from the original spectrum.}
	\label{fig:a_235_2d}
\end{figure}

\begin{figure}
	\includegraphics[angle=0,width=3.5in]{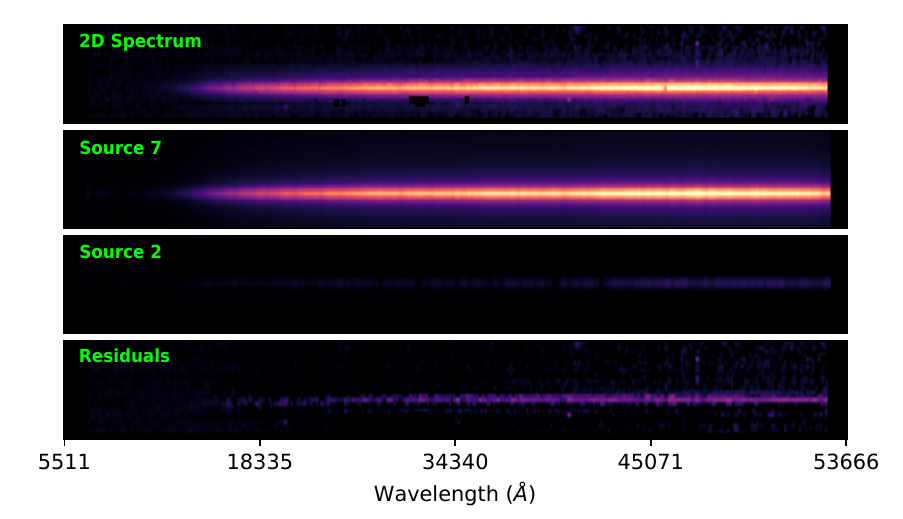}
	\caption{PRISM spectrum from Slit A. Top panel shows the 2D spectrum from the {\it JWST} pipeline. The middle two panels show the best-fit models of the host galaxy (Source 7) and the SN (Source 2). The bottom panel shows the residuals after subtracting the galaxy and the SN models from the original spectrum.}
	\label{fig:a_prism_2d}
\end{figure}

\begin{figure}
	\includegraphics[angle=0,width=3.5in]{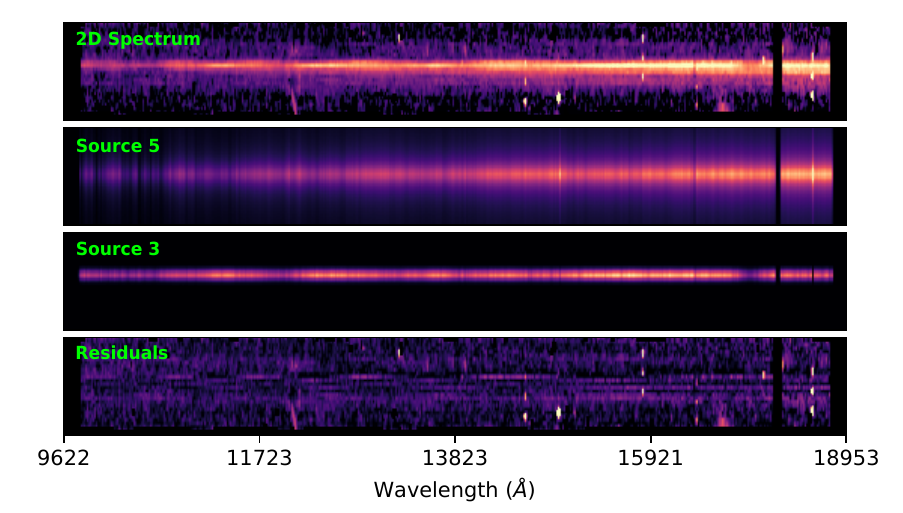}
	\caption{G140M spectrum from Slit B. Top panel shows the 2D spectrum from the {\it JWST} pipeline. The middle two panels show the best-fit models of the host galaxy (Source 5) and the SN (Source 3). The bottom panel shows the residuals after subtracting the galaxy and the SN models from the original spectrum.}
	\label{fig:b_140_2d}
\end{figure}

\begin{figure}
	\includegraphics[angle=0,width=3.5in]{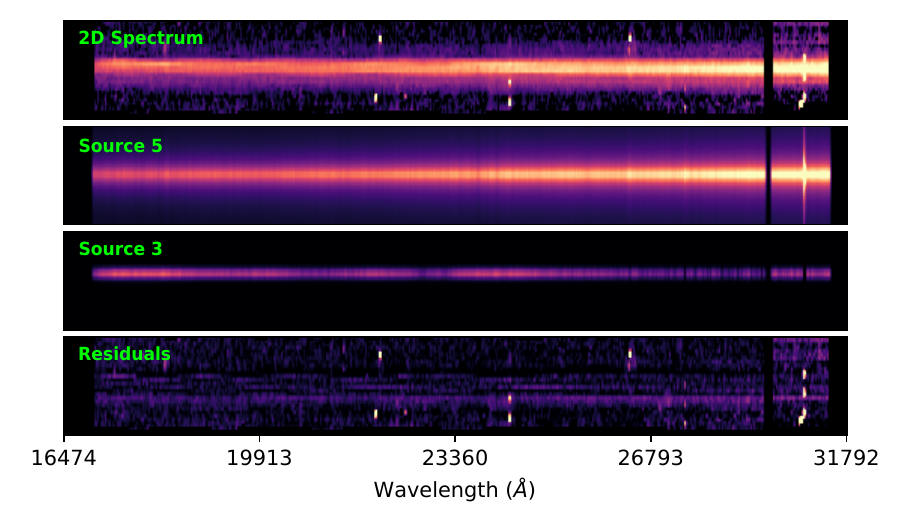}
	\caption{G235M spectrum from Slit B. Top panel shows the 2D spectrum from the {\it JWST} pipeline. The middle two panels show the best-fit models of the host galaxy (Source 5) and the SN (Source 3). The bottom panel shows the residuals after subtracting the galaxy and the SN models from the original spectrum.}
	\label{fig:b_235_2d}
\end{figure}

\begin{figure}
	\includegraphics[angle=0,width=3.5in]{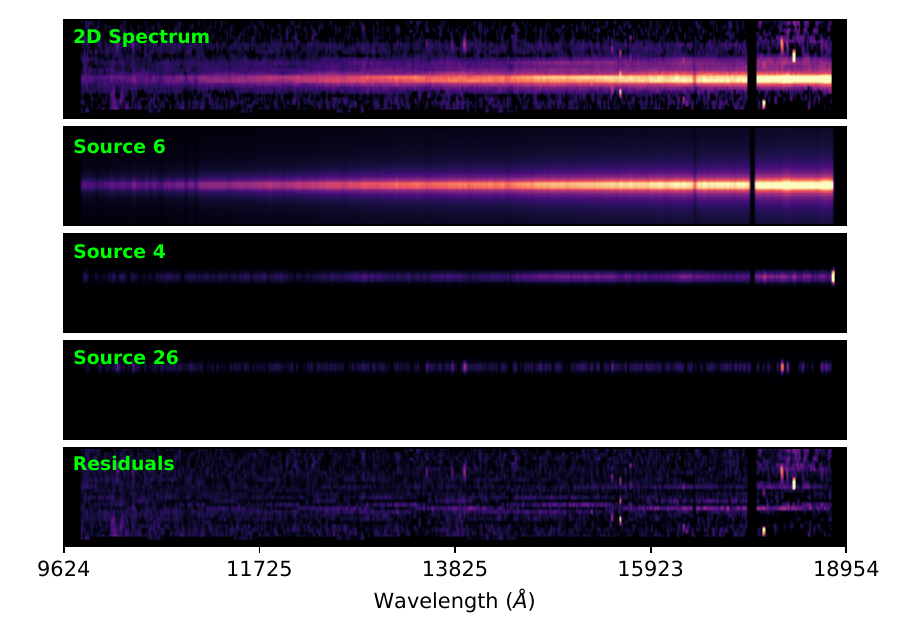}
	\caption{G140M spectrum from Slit C. Top panel shows the 2D spectrum from the {\it JWST} pipeline. The middle three panels show the best-fit models of the host galaxy (Source 6), the SN (Source 4), and a nearby arc (Source 26). The bottom panel shows the residuals after subtracting the galaxy and the SN models from the original spectrum.}
	\label{fig:c_140_2d}
\end{figure}

\begin{figure}
	\includegraphics[angle=0,width=3.5in]{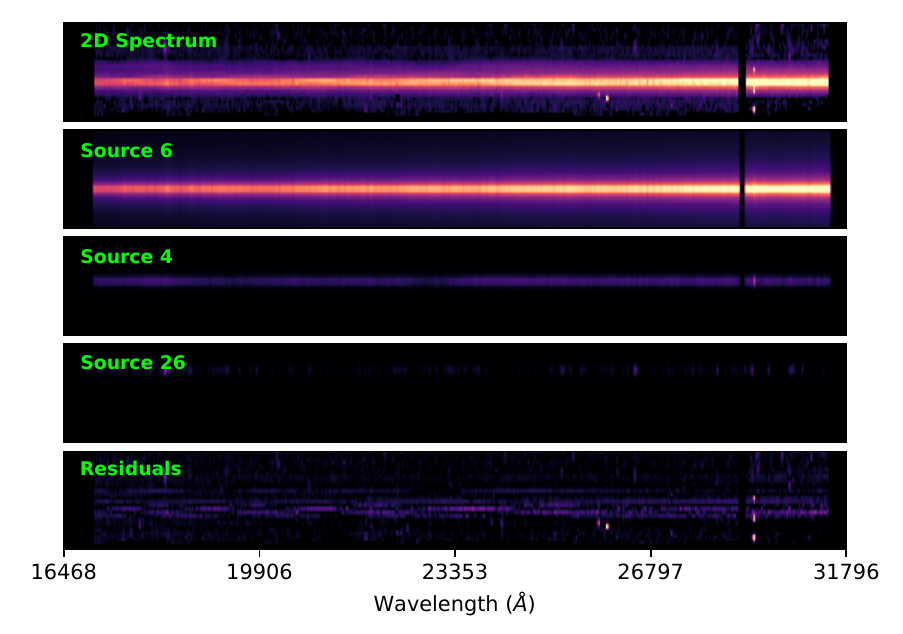}
	\caption{G235M spectrum from Slit C. Top panel shows the 2D spectrum from the {\it JWST} pipeline. The middle three panels show the best-fit models of the host galaxy (Source 6), the SN (Source 4), and a nearby arc (Source 26). The bottom panel shows the residuals after subtracting the galaxy and the SN models from the original spectrum.}
	\label{fig:c_235_2d}
\end{figure}

\begin{figure}
	\includegraphics[angle=0,width=3.5in]{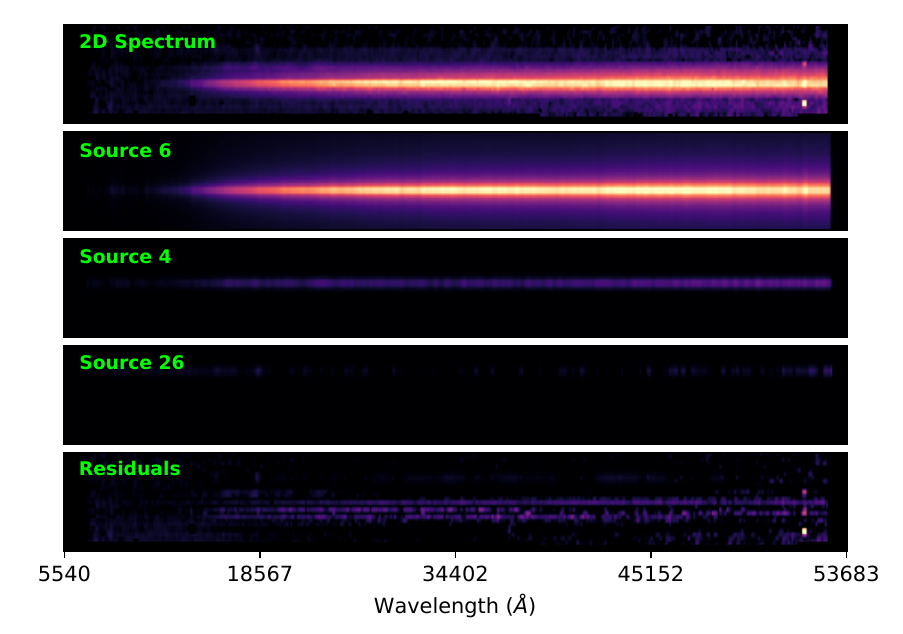}
	\caption{PRISM spectrum from Slit C. Top panel shows the 2D spectrum from the {\it JWST} pipeline. The middle three panels show the best-fit models of the host galaxy (Source 6), the SN (Source 4), and a nearby arc (Source 26). The bottom panel shows the residuals after subtracting the galaxy and the SN model from the original spectrum.}
	\label{fig:c_prism_2d}
\end{figure}

All the 2D spectra taken from Slits A, B, and C are shown in Figures~\ref{fig:a_140_2d} through~\ref{fig:c_prism_2d} (except the NIRSpec PRISM spectra from Slit B, which is shown in Figure~\ref{fig:2d_model}). Labels are source names in the MSA configuration data, where Sources 2, 3, and 4 are SN Images A, B, and C, respectively, and Sources 5, 6, and 7 are images of the SN's host galaxy for Images B, C, and A, respectively. 


\section{Results of the SN-Phase measurements before accounting for systematic errors}
\label{appendix:sn_phase_results}

Table~\ref{tab:td_results} presents the constraints on the phases of the SN images based on the analyses described in  Section~\ref{sec:sn_phases}. These constraints arise either from the deviation of the model parameters within the MCMC chains or from the variation in the templates used by the {\tt SNID} software package. The systematic uncertainties derived from our simulations, as described in Section~\ref{sec:method_validation}, are not included in the error estimates listed in Table~\ref{tab:td_results}.

\begin{table*}
\centering
\caption{\textbf{SN H0pe Image Phases}} 
\begin{tabular}{lccc}
\hline\hline
 & \textbf{Image A} & \textbf{Image B} & \textbf{Image C} \\
\hline
\textbf{Hsiao} (joint) & ${\tabtamhsiaojt}_\tabtalhsiaojt^\tabtauhsiaojt$ & ${\tabtbmhsiaojt}_\tabtblhsiaojt^\tabtbuhsiaojt$ & ${\tabtcmhsiaojt}_\tabtclhsiaojt^\tabtcuhsiaojt$ \\
\textbf{Hsiao} (independent) & ${\tabtamhsiao}_\tabtalhsiao^\tabtauhsiao$ & ${\tabtbmhsiao}_\tabtblhsiao^\tabtbuhsiao$ & ${\tabtcmhsiao}_\tabtclhsiao^\tabtcuhsiao$ \\
\textbf{SALT3-NIR} (joint) & ${\tabtam}_\tabtal^\tabtau$ & ${\tabtbm}_\tabtbl^\tabtbu$ & ${\tabtcm}_\tabtcl^\tabtcu$ \\
\textbf{SALT3-NIR} (independent) & ${\tabtamind}_\tabtalind^\tabtauind$ & ${\tabtbmind}_\tabtblind^\tabtbuind$ & ${\tabtcmind}_\tabtclind^\tabtcuind$ \\
\textbf{SNID} (templates 2.0) & \nodata & $\tabtbmsnida_{-\tabtbesnida}^{+\tabtbesnida}$ & $\tabtcmsnida_{-\tabtcesnida}^{+\tabtcesnida}$ \\
\spw{\textbf{SNID} (Ia-norm templates)} & \nodata & \ttbsnidsubtype & \ttcsnidsubtype \\
\textbf{SNID} (SALT3-NIR spectra) & \nodata & $\tabtbmsnidb_{-\tabtbesnidb}^{+\tabtbesnidb}$ & $\tabtcmsnidb_{-\tabtcesnidb}^{+\tabtcesnidb}$ \\
\hline
\end{tabular}
\tablecomments{Phases are rest-frame days after the peak brightness, and uncertainties are 1$\sigma$. For fits to the Hsiao07 templates and the SALT-NIR model, the listed uncertainties encompass solely the dispersion within their respective MCMC samples.}
\label{tab:td_results}
\end{table*}

\section{Results from the SALT3-NIR model}
\label{appendix:salt3nir_results} 

\subsection{Joint MCMC fitting}

\begin{figure*}
\centering
\includegraphics[angle=0,width=6.2in]{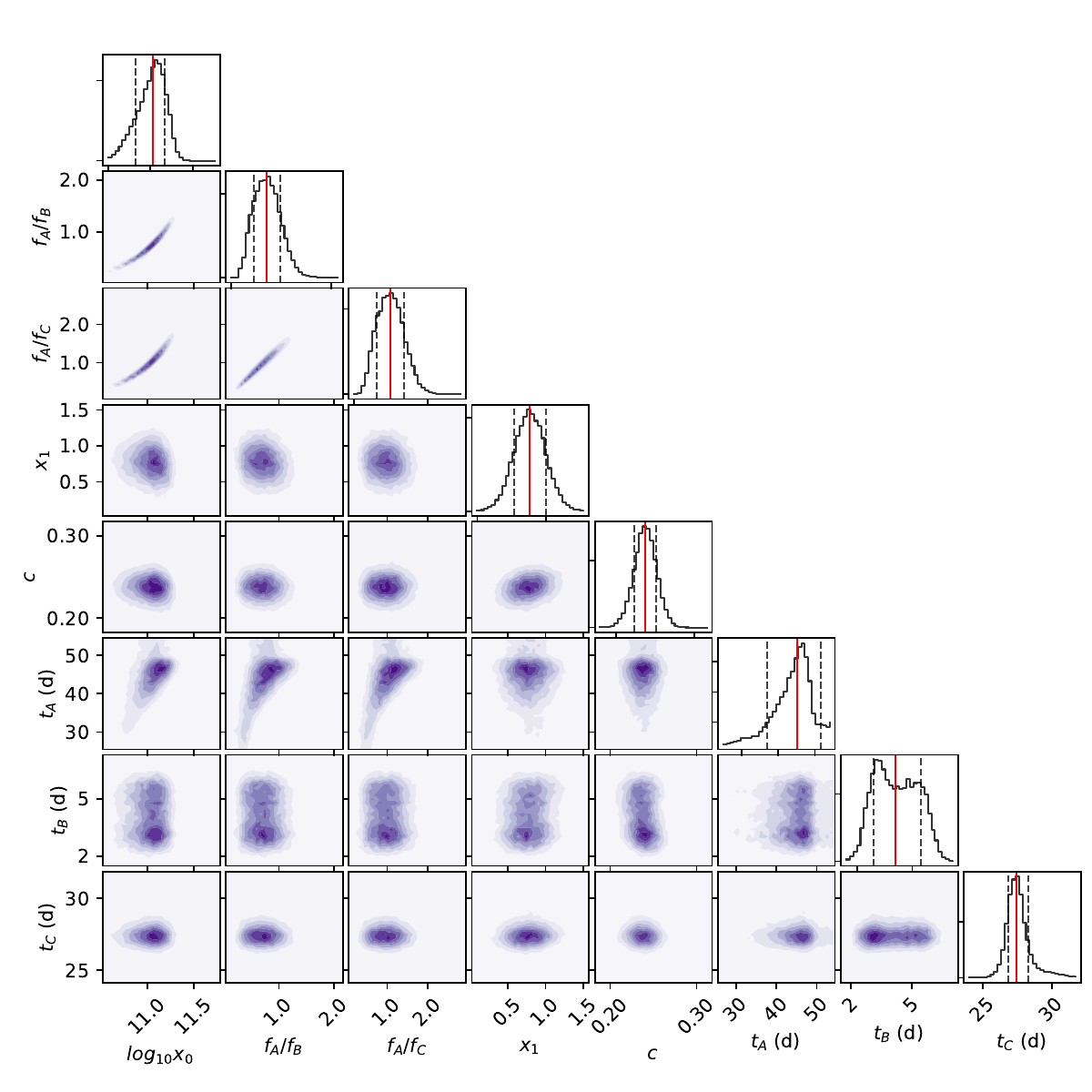}
\caption{Parameter distributions from the SALT3-NIR model fits.  $x_0$, $x_1$ and $c$ are SALT3-NIR model parameters, $f_A/f_B$ and $f_A/f_C$ are the template-flux ratios between Images A and B and between Images A and C, respectively, and $t_A$, $t_B$, and $t_C$ are rest-frame phases of Images A, B, and C, respectively. The solid vertical line on each histogram marks the median of the parameter, while the dashed vertical lines denote the 68\% confidence interval of each distribution.}
\label{fig:cornerplot_salt}
\end{figure*}

\begin{figure*}
\centering
\includegraphics[angle=0,width=5in]{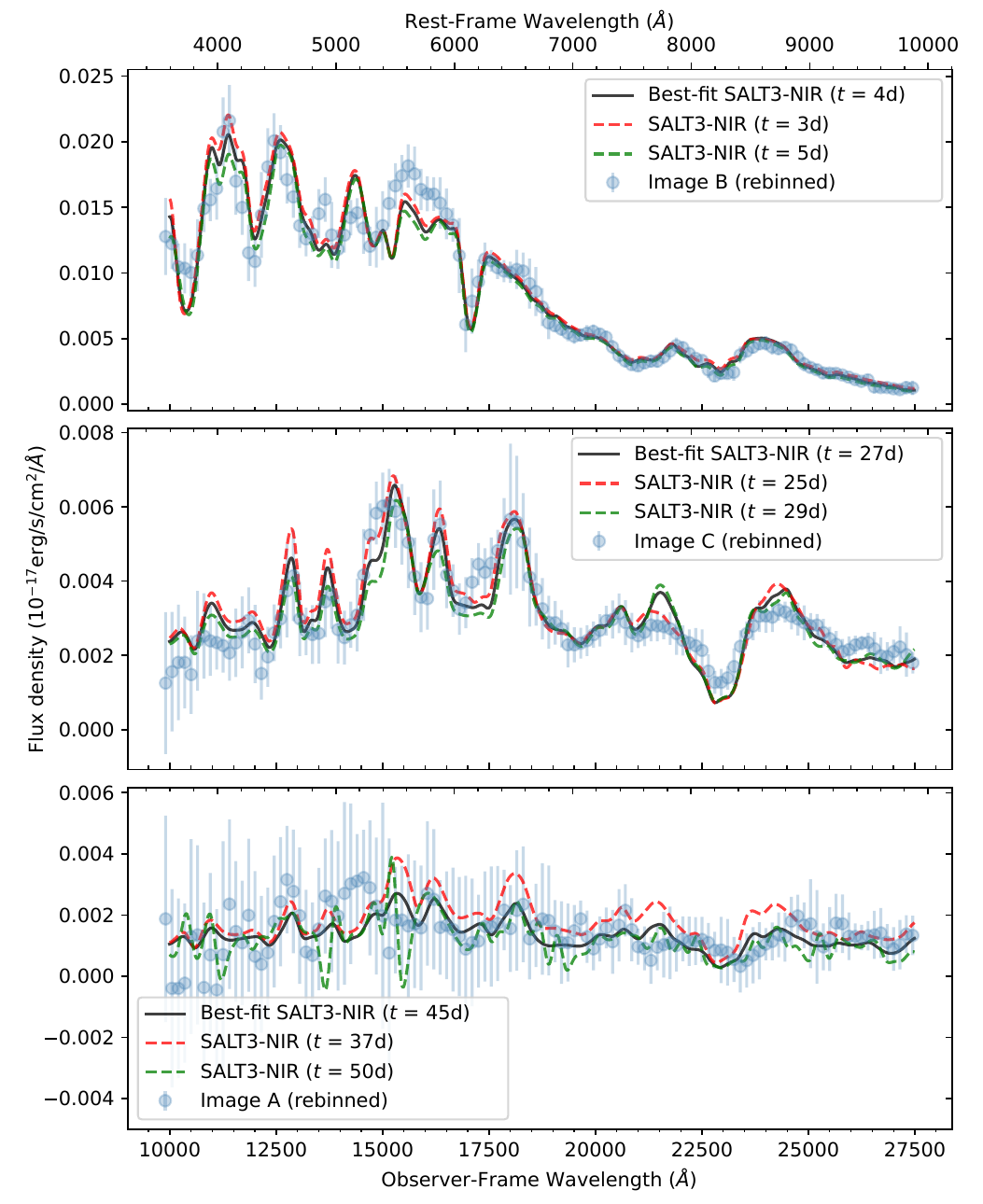}
\caption{Best-fit SALT3-NIR models from Figure~\ref{fig:snspec} are depicted again here, including the models for SN phases that deviate by $\pm 1\sigma$ (rounded to the nearest integers) from the best-fit phases. Data points and error bars show the observed (120\AA-rebinned) SN~H0pe spectra.}
\label{fig:salt_fitting_1sigma_mcmc}
\end{figure*}

The primary SALT3-NIR fit was simultaneous to all three images of SN H0pe. Shared parameters comprised spectral stretch $x_1$ and color $c$, as described in Section~\ref{sec:salt_fitting}. 
Figure~\ref{fig:cornerplot_salt} shows the posterior distributions of the fitted parameters. The fits constrain $x_1=0.77\pm0.23$ and  $c=0.24\pm0.02$.
Figure~\ref{fig:salt_fitting_1sigma_mcmc} illustrates the fit uncertainties. Absorption features at rest-frame $\sim$4000\,\AA, $\sim$4900\,\AA, and $\sim$5600\,\AA\ in the modeled spectrum at $t=50$ days (shown in the bottom panel of Figure~\ref{fig:salt_fitting_1sigma_mcmc}) are likely artifacts due to the model approaching its 50-day phase boundary. No such features have been observed in previous SN Ia spectra.

\subsection{Simulation of the model fitting}

\begin{figure*}
\centering
\includegraphics[angle=0,width=5in]{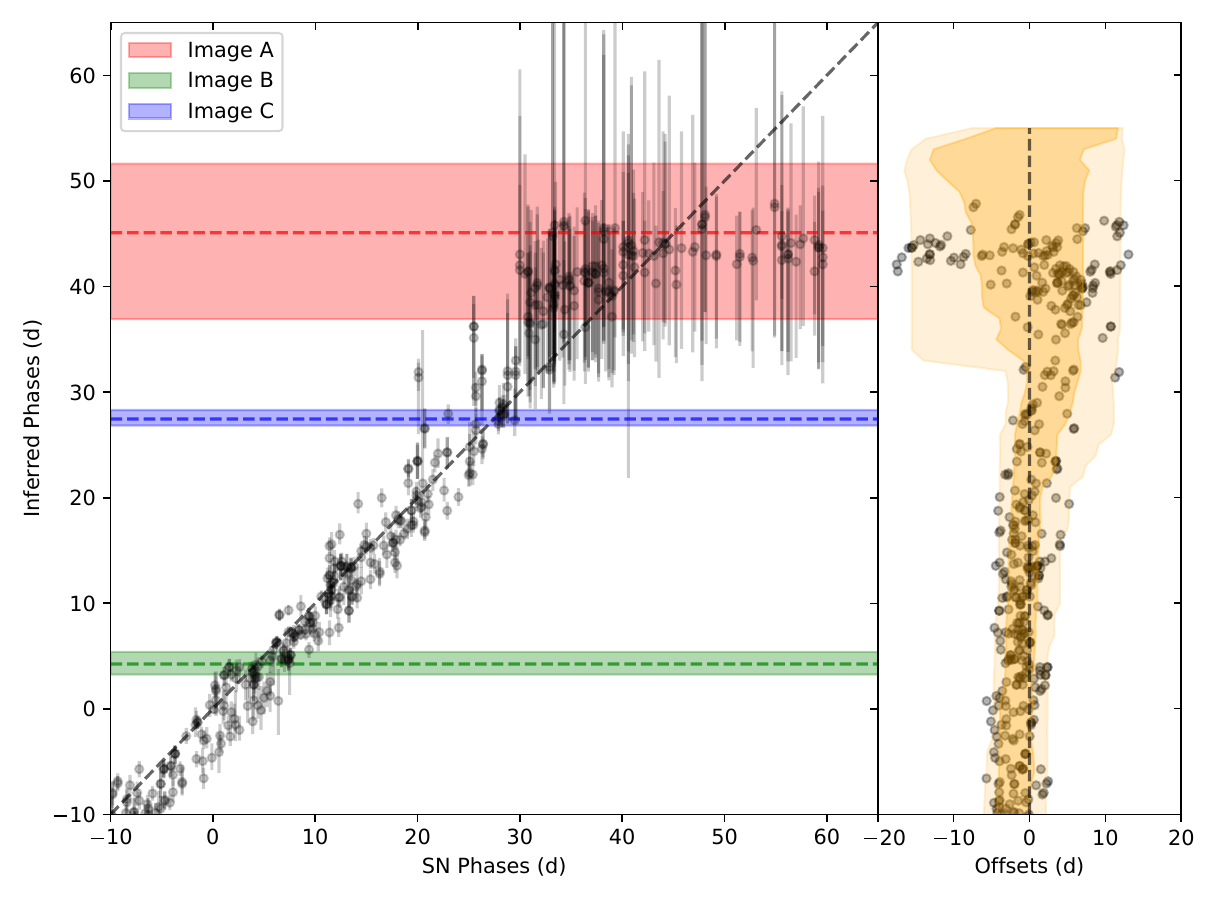}
\caption{{\bf Left:} inferred phases from SALT3-NIR fitting versus actual phases of a set of nearby SNe Ia. Error bars correspond to the 68\% confidence intervals of the inferred phases. Horizontal dashed lines and shaded regions mark the inferred phases and their 68\% confidence intervals for the three images of SN H0pe. The diagonal dashed line shows equality. {\bf Right:}  deviation of the best-fit inferred phase  from the actual phase. Shaded regions mark the 68\% (darker) and 95\% (lighter) confidence intervals of the residuals measured within a $10$-day rolling window.
\label{fig:validation_salt}}
\end{figure*}

We tested the SALT3-NIR results using simulations similar to those outlined in Section~\ref{sec:method_validation}. Figure~\ref{fig:validation_salt} shows inferred versus  actual phases for joint fits to a set of nearby SNe Ia.
\spw{The uncertainties are about the same as those of the Hsiao07 fits for phases $\la$30~days, and systematic offsets in that range are negligible.  At later phases,  the inferred phases tend to be a little later than the true phases but consistent with equality within the larger uncertainties.}

\subsection{Constraints on time delays and magnifications}
\label{sec:constraints_salt}

\begin{figure*}
	\centering
	\includegraphics[angle=0,width=5.5in]{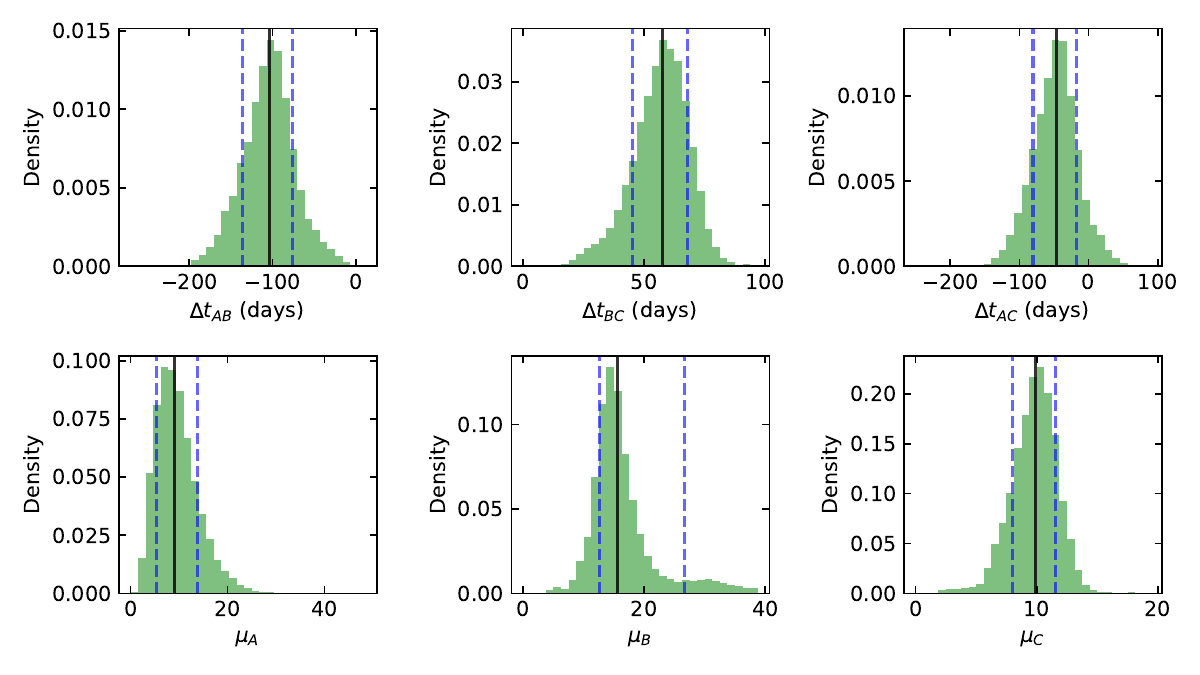}
	\caption{Posterior probability densities of the relative time delays and the macro lens' magnifications between SN H0pe's images. Histogram shows values derived from the SALT3-NIR model. The distributions integrate the uncertainty from the MCMC fitting with systematic uncertainties, estimated from the model fitting of nearby SN Ia spectra and from the simulated SN Ia spectra. The uncertainties take into account the impacts of microlensing and millilensing effects.}
	\label{fig:pstr_salt}
\end{figure*}

We repeated the Section~\ref{sec:final_constraints} analysis using the SALT3-NIR model, fixing the spectral stretch and color parameters $x_1$ and $c$ to the best-fit values.
\spw{The resulting constraints on the phases $t_i$, time delays $\Delta t_i$, and magnifications $\mu_i$ are shown in Table~\ref{tab:final_result}. Values are consistent with those from the Hsiao07 fits within the uncertainties.}
%
\spw{Calculating absolute magnifications is possible because}
the luminosities of SNe Ia show small dispersion after correction for light-curve shape and color. For the SALT3-NIR model, we corrected their $B$-band magnitudes for the spectral stretch ($x_1$) and color ($c$) based on the Tripp equation \citep{1998A&A...331..815T}, and hence the corrected $B$-band magnitude ($\hat{m}_B$) is given by $\hat{m}_B=\mu + M=m_B+\alpha x_1-\beta c$, where $\mu$ is the distance modulus, $M$ is the absolute $B$-band magnitude of a fiducial SN Ia with $x_1=0$ and $c=0$, and $m_B$ is the measured apparent peak magnitude in rest-frame $B$-band. We used $\alpha=0.15$ and $\beta=3.0$ from \citet{scolnic2018} based on the Gaussian intrinsic scatter model of SNe Ia from \citet{2010A&A...523A...7G}. 
Subsequently, comparing $\hat{m}_B$ from peak SALT3-NIR spectra derived from a set of $\mathbf{Y}$ parameters from the MCMC chain to the expected $\hat{m}_B$ based on the complete sample from \citet{scolnic2018} gave the posterior distributions of the magnifications listed in Table~\ref{tab:final_result}.
An alternative assumption is that SN H0pe is a standard candle in the rest-frame $Y$-band, as described in Section~\ref{sec:final_constraints}. The $Y$-band magnitude of a standard SN Ia is characterized by $x_1=0$ and $c=0$ and a rest-frame $B$-band magnitude equal to the expected $\hat{m}_B$. 
This gave results consistent with the $B$-band magnitudes, after applying corrections for spectral stretch and color.
The posterior distributions of the SALT3-NIR magnifications are shown in Figure~\ref{fig:pstr_salt}, and  Table~\ref{tab:correlation_hsiao} lists the correlation matrices.

\end{document}